\documentclass[aps,prb,onecolumn,showpacs,showkeys,amsmath,amssymb,byrevtex]{revtex4}
\newcommand{\we}[1]{\mathbf{#1}}
\begin{document}
\title{Extension of Fr\"ohlich's method to 4-fermion interactions}
\author{Przemys\l aw \surname{Tarasewicz}}
\email[]{tarasek1@amb.bydgoszcz.pl}
\affiliation{Uniwersytet M.~Kopernika, Collegium Medicum, Wydzia\l\ Farmacji, 
Jagiello\'nska 13--15, 85--067 Bydgoszcz, Poland}
\author{Dominik \surname{Baran}}
\email[]{faramir@phys.uni.torun.pl}
\affiliation{Uniwersytet M.~Kopernika, Instytut Fizyki, Grudzi\c adzka 5, 87--100 
Toru\'n, Poland}
\date{\today}
\begin{abstract}
Higher order terms of the~transformed electron-phonon Hamiltonian~$H_{e-ph}$, 
obtained by performing the~Fr\"ohlich's~transformation, are investigated. 
The~influence of~terms discarded by~Fr\"ohlich (in particular those proportional to 
the~third power of electron-phonon coupling) on the~effective Hamiltonian is 
examined. To this end a~second Fr\"ohlich-type transformation is performed, which 
yields, among others, an~effective 4-electron interaction. This interaction is 
reduced to a~form admitting solution of~thermodynamics. The form of the~coupling of 
the~4-electron interaction is found. By applying standard approximations, it is 
shown that this interaction is attractive with interaction coupling given 
by~$-D_{\we{k}_F}^6 / \omega_{\we{k}_F}^5$, where $D_{\we{k}}$ is electron-phonon 
coupling, $\omega_{\we{k}}$ is phonon energy and $\we{k}_F$ is Fermi momentum.

The form of higher order terms of the~original Fr\"ohlich-transformed~$H_{e-ph}$ 
are also found, up to terms proportional to the~6-th power of the~coupling, that is 
up to those, which yield the~effective 4-electron interactions.
\end{abstract}
\maketitle
\section{Introduction}
The~BCS theory\cite{bcs} proved extremely effective as a~theory of superconductivity. 
The~key idea of this theory---the~attractive interaction which binds electrons into 
Cooper pairs---has its essential origins in earlier conclusions drawn 
Fr\"ohlich\cite{fr:1}, who performed a~unitary transformation~$U$ of 
the~electron-phonon Hamiltonian~$H_{e-ph}$. $U$~was adjusted so as to eliminate 
the~electron-phonon interaction as far as possible and replace it by an~effective 
interaction~$V_{eff}$ between electrons dressed in the phonon field. 
$V_{eff}$~proved to be attractive for 1-electron energies close to~$\varepsilon_F$. 
A~reduced form of~$V_{eff}$ was subsequently used by~Bardeen, Cooper and~Schrieffer 
in their theory\cite{bcs}.

It is worth emphasizing that Fr\"ohlich's~transformation is not~strictly unitary, 
because the~less significant terms of the~resulting expansion of~$UH_{e-ph}U^{-1}$ 
were discarded. Correctness of the~remaining terms, included into the~Hamiltonian of
a~superconductor, was confirmed by the~success of the~BCS theory.

Unfortunately, BCS~theory proved incapable of explaining superconductivity in 
type-II superconductors, heavy fermions and high-$T_c$ superconductors~(HTSC). 
The~search for an~alternative theory of superconductivity proceeds in various 
directions and one of them exploits the~idea of extending BCS~theory by adding 
to the~BCS Hamiltonian~$H_{BCS}$ further interactions. Rickayzen\cite{rick} 
suggested incorporation of a~4-fermion interaction, motivating such choice by 
analogies between theory of superconductivity and nuclear theory, where such 
interactions had been considered. This idea was also mentioned 
by~Volovik\cite{volovik}, but remained in the~realm of theoretical concepts until 
the~late nineties, when Ma\'ckowiak and one of 
authors~(PT)\cite{mt:3,mt:4,mt:2,mt:1,tm,t:1,t:2} (MT) proposed a 
Hamiltonian~$H_{MT}=H_{BCS}+W+V_{MT}$, where
\begin{equation}
	H_{BCS}=T+V_{BCS}=\sum_{\we{k}\sigma}\xi_{\we{k}}n_{\we{k}\sigma}
	-|\Lambda|^{-1}\sum_{\we{k}\we{k'}}G_{\we{k}\we{k'}}a_{\we{k}+}^\ast 
	a_{-\we{k}-}^\ast a_{-\we{k'}-}a_{\we{k'}+},
\label{H_BCS}
\end{equation}
\begin{equation}
	W=\sum_{\we{k}}\gamma_{\we{k}}n_{\we{k}+}n_{\we{k}-},
\label{W}
\end{equation}
\begin{equation}
	V_{MT}=|\Lambda|^{-1}\sum_{\we{k},\we{k'}}g_{\we{k}\we{k'}}
	a_{\we{k}-}^\ast a_{\we{k}+}^\ast a_{-\we{k}-}^\ast a_{-\we{k}+}^\ast 
	a_{-\we{k'}+}a_{-\we{k'}-}a_{\we{k'}+}a_{\we{k'}-},
\label{V}
\end{equation}
$|\Lambda|$ in Eqs.~\eqref{H_BCS}--\eqref{V} denotes the system's volume, 
whereas~$g_{\we{k}\we{k'}}$, $G_{\we{k}\we{k'}}$ are bounded functions 
and~$\gamma_{\we{k}}$, $g_{\we{k}\we{k'}}$, $G_{\we{k}\we{k'}}$ are invariant under 
time reversal~$\we{k}\to -\we{k}$ or~$\we{k'}\to-\we{k'}$.

Considerations which led to the~introduction of~$W$ were founded on the~analysis of 
the~HTSC normal state. There are grounds to believe that this is not a~normal 
Fermi~liquid state. The interaction~$W$, first added to~$H_{BCS}$ 
by~Czerwonko\cite{cz:1,cz:2}, guaranteed normal-state behaviour characteristic of 
the~so-called statistical spin liquid, considered earlier by~Spa\l ek 
and~\mbox{W\'ojcik}\cite{spawoj:1, spawoj:2}.

$W$ is a~2-electron interaction. This raises the~question, whether it can be 
obtained by a~reduction procedure (different than the BCS~one) of the~interaction 
derived by~Fr\"ohlich. It can be easily verified that this is impossible. More 
precisely, for the~unique possible reduction of~momenta, the~coupling vanishes, 
meaning that the~nature of~$W$ is not~phononic.

Introduction of the~4-fermion interaction~$V_{MT}$ was justified in 
Ref.~\onlinecite{mt:3} by its possible role as an~attraction between pairs in~HTSC, 
mediated by phonons or other quanta\cite{mt:1}. A~conjecture put forward in 
Ref.~\onlinecite{mt:3} suggested also that~$V_{MT}$ could be expected to arise as 
one of the~higher-order terms of Fr\"ohlich's~expansion of~$UH_{e-ph}U^{-1}$. 
An~alternative justification was given in Ref.~\onlinecite{mt:1}, where~$V_{MT}$ 
was viewed as a~BCS-type interaction between quasi-particles of a~free gas 
represented by~$W$ written in the~form~$\sum_{\we{k}}\gamma_{\we{k}}c_{\we{k}}^\ast 
c_{\we{k}}$, with~$c_{\we{k}}=a_{\we{k}+}a_{\we{k}-}$. Both of these ideas 
essentially exploit the~concept of phonon-type mediation of interactions. 
The~significance of this mediation in HTSC was stressed 
by~Wysoki\'nski~\cite{wysok:1}.

The~question of the~form of higher-order terms of Fr\"ohlich's expansion is 
interesting itself, not only as providing a possible explanation of 
the~MT extension, but, first of all, because these terms could throw some light on 
further possible extensions of~$H_{BCS}$ related to effects of electron-lattice 
interaction. Since Fermi-liquid theory will remain the~foundation of our formalism, 
we shall focus our interest on effective electron interactions.

It is worth noting that the~possible presence of fermion quadruples in superconductors
and superfluids was considered in a~number of papers. Schneider and~Keller\cite{schn:1}
measured the~various characteristics of some cuprates and Chevrel-phase superconductors,
especially concentrating on the~relation between the~critical temperature and zero 
temperature condensate density. They noticed that the~experimental data for 
e.g.,~$\mathrm{Y}\mathrm{Ba}_2 \mathrm{Cu}_3 \mathrm{O}_{6.602}$ point to similarities
with the~behaviour of a~dilute Bose~gas. As a~result they suggested Bose condensation of 
weakly interacting fermion pairs as a~mechanism of transition from normal to 
superconducting state. Bunkov~et~al.\cite{bunkov} pointed to presence of fermion 
quadruples in~$\mbox{}^3\mathrm{He}$. Their work was devoted to the~problem of 
influence of spatial disorder on the~order parameter in 
superfluid~$\mbox{}^3\mathrm{He}$. By resorting to the~work of~Volovik\cite{volovik}, they suggested that unusual spectra of~$\mbox{}^3\mathrm{He}$ in aerogel could be 
explained by a~process in which impurities tend to destroy the~anisotropic 
correlations of the~order parameter, while correlations of higher symmetry can survive 
(e.g.,~four-particle correlations). Recently Schneider~et~al.\cite{schn:2} discovered 
of~half-$h/2e$ magnetic flux in SQUIDs  fabricated of 
bicrystalline~$\mathrm{Y}\mathrm{Ba}_2 \mathrm{Cu}_3 \mathrm{O}_{7-\delta}$ films. 
This situation corresponds to the~presence of fermion quadruples in the~system. 
Based on this observation, Aligia~et~al.\cite{aligia} investigated a~model of interface 
between two superconductors, based on a~one~dimensional boson lattice model and
proposed formation of quartets of electrons.

In Section~\ref{2} higher order terms of the~expansion of 
Fr\"ohlich's~transformation~$UH_{e-ph}U^{-1}$ are discussed qualitatively in order to 
exhibit the~emerging structure. Owing to the~complexity of this procedure, only terms 
proportional to the~third power of the~electron-phonon coupling are found in 
Section~\ref{3}. The~resulting extended Fr\"ohlich Hamiltonian is transformed in the 
next section by a~second Fr\"ohlich-type transformation, which produces 4-electron 
terms. These are discussed in detail in Section~\ref{5}. In particular, a~reduction 
similar to the~BCS one is performed, which yields an~interaction 
of the~form~$V_{MT}$ in Eq.~\eqref{V}. The~expression for~$g_{\we{k}\we{k'}}$ is 
derived. Detailed analysis of this expression is performed in the~next section, in 
particular, we showed, by applying some approximations, that this expression is 
negative i.e.,~$V_{MT}$ is attractive. The~original Fr\"ohlich's~transformation and 
the~resulting terms, up to 6th power of the~coupling, are commented in 
Section~\ref{6}. The final section contains a~discussion, summary and open questions.
\section{\label{2}Higher order terms of the Fr\"ohlich's transformation}
Following Fr\"ohlich\cite{fr:1} (see Appendix~\ref{a} for details), let us consider
the~electron-phonon Hamiltonian:
\begin{equation}
	H_{e-ph}=H_0+H_{int}=\sum_{\we{k}\sigma}\varepsilon_{\we{k}}
	a_{\we{k}\sigma}^\ast a_{\we{k}\sigma}+
	\sum_{\we{w}}\omega_{\we{w}}b_{\we{w}}^\ast b_{\we{w}}+
	i\sum_{\we{w}}D_{\we{w}}\left( 
  	b_{\we{w}}\rho_{\we{w}}^\ast -b_{\we{w}}^\ast \rho_{\we{w}}\right),
\label{H_e-ph}
\end{equation}
where
\begin{equation}
  \rho_{\we{w}}=\sum_{\we{k}\sigma}a_{\we{k}-\we{w}\sigma}^\ast 
  a_{\we{k}\sigma},
\label{rho}
\end{equation}
and~$a_{\we{k}\sigma}$ ($b_{\we{k}}$) are fermion (boson) operators. 
The~coupling~$D_{\we{w}}$ will be assumed small and~$\hslash\equiv 1$. 

Since the~interaction is spin-independent, the spin index will be suppressed. 
Summation over electron momenta will include summation over spins.

Fr\"ohlich performed a~unitary transformation of~$H_{e-ph}$ in order to eliminate 
(as far as possible) the~interaction term. The~transformed Hamiltonian is
\begin{equation}
	H=\mathrm{e}^{S^\ast}H_{\mathrm{e-ph}}\mathrm{e}^S=H_{\mathrm{e-ph}}-
	[S,H_{\mathrm{e-ph}}]+
        \frac{1}{2}[S,[S,H_{\mathrm{e-ph}}]]+\ldots\, ,
\label{h_trans}
\end{equation}
where
\begin{equation}
	S=\sum_{\we{q}}S_{\we{q}}=\sum_{\we{q}}\left( \gamma_{\we{q}}^\ast 
	b_{\we{q}}^\ast -\gamma_{\we{q}} b_{\we{q}}\right)=-S^\ast, 
\label{S}
\end{equation}
\begin{equation}
	\gamma_{\we{q}}=\sum_{\we{k}}\phi(\we{k},\we{q})a_{\we{k}}^\ast 
	a_{\we{k}-\we{q}},
\label{gamma}
\end{equation}
and the~unknown function~$\phi(\we{k},\we{q}): 
\mathbb{R}^3\times\mathbb{R}^3\to\mathbb{C}^1$ is adjusted to achieve the~cancellation.
		
Subsequently, a~term which is a~combination of products, each with $f$~fermion operators 
and $b$~boson operators will be written as~$(f,b)$. Clearly, $f$~will always be even. 
For example, $H_0$~consists of terms~$(2,0)$ and~$(0,2)$. 

The~rhs of Eq.~\eqref{h_trans} expresses in terms of commutators~$[(f_1,b_1),(f_2,b_2)]$. 
One easily finds that 
\begin{equation}
	[(f_1,b_1),(f_2,b_2)]=[f_1,f_2]b_1 b_2 + f_2 f_1 [b_1,b_2]=[f_1,f_2]b_2 b_1 + 
	f_1 f_2 [b_1,b_2].
\label{com}
\end{equation}
The necessary commutators~$[f_1,f_2]$, $[b_1,b_2]$ are given in Appendix~\ref{b}.

According to Eq.~\eqref{h_trans}, the~transformation can be performed, given commutators 
of the~form occurring in Eq.~\eqref{com} with the~first argument equal~$S$. The latter 
is a~$(2,1)$~expression, hence 
\begin{equation}
  [S,(f,b)]=[(2,1),(f,b)]=(f,b+1)+(f+2,b-1),
\label{com_s}
\end{equation}
by virtue of Eqs.~\eqref{com},~\eqref{com_4f}. Clearly~$(f,b-1)=0$ for~$b=0$. 

Based on this ground, Fr\"ohlich obtained the~transformed Hamiltonian 
(see Appendix~\ref{a}):
\begin{equation*}
	H_F=\sum_{\we{k}}\varepsilon_{\we{k}}n_{\we{k}}-
	\frac{1}{2} \sum_{\we{k}\we{q}\we{w}} 
	\frac{D_{\we{w}}^2 \left( 1+\Delta\left(\we{k},\we{w}\right)\right) 
	\left( 1-\Delta(\we{q},\we{w})\right)}{\varepsilon_{\we{q}-\we{w}}-
	\varepsilon_{\we{q}}+\omega_{\we{w}}}\left( a_{\we{k}}^\ast a_{\we{k}-
	\we{w}}a_{\we{q}-\we{w}}^\ast a_{\we{q}} +c.c.\right).
\end{equation*}
The second term represents an~effective interaction between electrons dressed in 
the~phonon field. If~$\varepsilon_{\we{q}-\we{w}}-\varepsilon_{\we{q}}+
\omega_{\we{w}}>0$, this interaction is attractive.

Omission of higher terms in Eq.~\eqref{h_trans} results in violation of unitarity. 
The~question thus arises whether partial inclusion of these terms (first of all 
those of 3rd~order in the~coupling) could improve agreement between theory and 
experiment.

Following the~general rule for the~action of~$S$ in consecutive orders, expressed 
by Eq.~\eqref{com_s}, one easily finds the~form of subsequent terms:
\begin{equation}
	\left\{ 
		\begin{aligned} (2,0) \\ (0,2) \end{aligned} 
	\right\} \xrightarrow{S} (2,1) \xrightarrow{S} 
	\left\{ 
		\begin{aligned} (4,0) \\ (2,2) \end{aligned} 
	\right\} \xrightarrow{S} (4,1) \xrightarrow{S}
	\left\{ 
		\begin{aligned} (6,0)\\(4,2) \end{aligned} 
	\right\} \xrightarrow{S} (6,1) \xrightarrow{S}
	\left\{ 
		\begin{aligned} (8,0) \\ (6,2) \end{aligned} 
	\right\}.
\label{sgraf}
\end{equation}
In each consecutive step one obtains terms proportional to the~next power of 
the~coupling.

From the~view-point of Fermi~liquid theory, the~terms representing effective 
inter-electron interactions are most interesting. According to Eq.~\eqref{sgraf}, 
new terms~$(6,0)$ (proportional to~$D_{\we{w}}^4$), describing 3-electron 
interactions and~$(8,0)$ (proportional to~$D_{\we{w}}^6$), representing 4-electron 
interactions, appear. Attractive interactions of this type leading to the~formation 
of fermion triples and quadruples, could affect the~behaviour of a~superconductor. 
However, the~total spin of an~electron triple is nonzero, so such clusters are 
unstable, as they are not invariant under time inversion. So far, there has been 
no experimental evidence of such objects.

Most electrons in a~superconductor below~$T_c$ are paired, so the 4-electron 
interaction between Cooper~pairs can be expected to prevail. Furthermore, quadruples 
with a~total spin equal zero and appropriate 1-electron momenta are stable under 
time inversion. On the~other hand, under Fr\"ohlich's~conditions for convergence of 
series \eqref{h_trans}, the~effect of the~terms~$(6,0)$,~$(8,0)$ is weaker.

Evaluation of these terms is a~complicated procedure. Before doing this, let us 
first examine the~3rd~order corrections.
\section{\label{3}Third order of the transformation}
Let us consider the~effect of the~first higher orders discarded by Fr\"ohlich, i.e.,
terms proportional to~$D_{\we{w}}^3$. Then the~corrected Hamiltonian takes the~form
\begin{equation}
	H'=H_0-\left(\left[S,H_0\right]-H_{int}\right)+
	\left(\frac{1}{2}\left[S,\left[S,H_0\right]\right] - 
	\left[ S,H_{int}\right]\right)
	-\left(\frac{1}{6}\left[S,\left[ S, \left[ S,H_0\right]\right]\right]
	-\frac{1}{2}\left[S,\left[S,H_{int}\right]\right]\right)+\ldots\, .
\label{trans_3rd_order}
\end{equation}
The~additional two terms in the~last bracket are equal, explicitly, 
\begin{equation}
\begin{split}
	& \frac{1}{2}\left[S,\left[ S,H_{int}\right]\right]-
	\frac{1}{6}\left[S,\left[ S, \left[ S, H_0 \right]\right] \right]
	=\sum_{\we{q}\we{k}}A_{\we{k}\we{q}}b_{\we{q}}^\ast 
	n_{\we{k}}\gamma_{\we{q}}^\ast\\
	& + \sum_{\we{q}\we{k}\we{w}\we{k'}}b_{\we{w}}^\ast \left\{ 
	B_{\we{k}\we{q}\we{w}\we{k'}} a_{\we{k}-\we{w}}^\ast a_{\we{k}-\we{q}}
	a_{\we{k'}-\we{q}}^\ast a_{\we{k'}}
	+ C_{\we{k}\we{q}\we{w}\we{k'}}a_{\we{k'}}^\ast a_{\we{k'}-\we{q}}
	a_{\we{k}-\we{q}}^\ast a_{\we{k}+ \we{w}}\right\}+c.c.:=H_e,\\	
\end{split}
\label{3order}
\end{equation}
where
\begin{equation}
	2A_{\we{k}\we{q}}=iD_{\we{q}}\phi^\ast(\we{k},\we{q})+
	iD_{\we{q}}\phi(\we{k}+\we{q},\we{q})
	+\frac{1}{3}\left(\varepsilon_{\we{k}-\we{q}}-\varepsilon_{\we{k}}+
	\omega_{\we{q}} \right) |\phi(\we{k},\we{q})|^2-\frac{1}{3}
	\left( \varepsilon_{\we{k}}-\varepsilon_{\we{k}+\we{q}}
	+\omega_{\we{q}}\right)|\phi(\we{k}+\we{q},\we{q})|^2 +c.c.,
\label{A}
\end{equation}
\begin{equation}
\begin{split}
	& 2B_{\we{k}\we{q}\we{w}\we{k'}} = iD_{\we{q}}\Bigl\{\phi^\ast(\we{k'},\we{q}) 
	\phi^\ast(\we{k-\we{q}},\we{w})-\phi^\ast(\we{k'},\we{q})
	\phi^\ast(\we{k},\we{w})
	+\phi^\ast(\we{k},\we{w})\phi(\we{k},\we{q}) -\phi^\ast(\we{k}-
	\we{q},\we{w})\phi(\we{k}-\we{w},\we{q})\Bigr\}\\
        &+\frac{1}{3}\phi^\ast(\we{k'},\we{q})\Bigl\{ \phi(\we{k},\we{q})
	\phi^\ast(\we{k},\we{w})\left(\varepsilon_{\we{k}}-\varepsilon_{\we{k}-
	\we{q}}+\varepsilon_{\we{k'}-\we{q}}-\varepsilon_{\we{k'}}\right)
	+\phi(\we{k}-\we{w},\we{q})\phi^\ast(\we{k}-\we{q},\we{w}) 
	\left(\varepsilon_{\we{k}-\we{w}-\we{q}}-\varepsilon_{\we{k}-\we{w}}-
	\varepsilon_{\we{k'}-\we{q}}+\varepsilon_{\we{k'}}\right)\Bigr\},\\
\end{split}
\label{B}
\end{equation}
\begin{equation}
\begin{split}
	2C_{\we{k}\we{q}\we{w}\we{k'}}&=
	iD_{\we{q}}\Bigl\{ \phi(\we{k'},\we{q})\phi^\ast(\we{k}+\we{w}
	-\we{q},\we{w})-\phi(\we{k'},\we{q})\phi^\ast(\we{k}+\we{w},\we{w})\\
	& - \phi^\ast(\we{k}+\we{w}-\we{q},\we{w})\phi^\ast(\we{k}+\we{w},\we{q})+
	\phi^\ast(\we{k}+\we{w},\we{w})\phi^\ast(\we{k},\we{q})\Bigr\}\\
	&+\frac{1}{3}\phi(\we{k'},\we{q})\Bigl\{ \phi^\ast(\we{k}+\we{w},\we{q}) 
	\phi^\ast(\we{k}+\we{w}-\we{q},\we{w}) \left( 
	\varepsilon_{\we{k'}-\we{q}}-\varepsilon_{\we{k'}}-
	\varepsilon_{\we{k}+\we{w}-\we{q}}+\varepsilon_{\we{k}+\we{w}}\right)\\
	& +\phi^\ast(\we{k},\we{q})\phi^\ast(\we{k}+\we{w},\we{w})
	\left( \varepsilon_{\we{k}-\we{q}}-\varepsilon_{\we{k}}-
	\varepsilon_{\we{k'}-\we{q}}+\varepsilon_{\we{k'}}\right)\Bigr\},\\
\label{C}
\end{split}
\end{equation}
It can be seen that a~new phonon index~$\we{w}$ has appeared in Eq.~\eqref{3order}. 
It results from the~commutator of~$S$ with the~terms~$(4,0)$, so the~harmonic 
approximation has not been violated.

Substitution of the~expression~\eqref{3order} into~$H'$ yields
\begin{equation*}
	H'=H_a+H_b+H_c+H_d+H_e,
\end{equation*}
where
\begin{equation*}
	H_a=\frac{1}{2}\sum_{\we{k}}\varepsilon_{\we{k}} n_{\we{k}}+
	\frac{1}{2}\sum_{\we{q}}\omega_{\we{q}}b_{\we{q}}^\ast b_{\we{q}}+c.c.,
\end{equation*}
\begin{equation}
	H_b=\sum_{\we{q}\we{k}}b_{\we{q}}^\ast b_{\we{q}}n_{\we{k}}
	\Bigl\{ iD_{\we{q}}\left(\phi(\we{k} +\we{q},\we{q})-
	\phi(\we{k},\we{q})\right)
	 -\frac{1}{2}\left(\varepsilon_{\we{k}}-\varepsilon_{\we{k}+\we{q}}+
	 \omega_{\we{q}}\right) |\phi(\we{k}+\we{q},\we{q})|^2+\frac{1}{2} 
	 \left(\varepsilon_{\we{k}-\we{q}} -\varepsilon_{\we{k}}+
	 \omega_{\we{q}}\right)|\phi(\we{k},\we{q})|^2 \Bigr\} +c.c.,
\label{H_b}
\end{equation}
\begin{equation}
	H_c=\sum_{\we{q}\we{k}}b_{\we{q}}\Bigl( iD_{\we{q}}+\left(
	\varepsilon_{\we{k}-\we{q}}-\varepsilon_{\we{k}}+\omega_{\we{q}} 
	\right)\phi(\we{k},\we{q})\Bigr) 
	a_{\we{k}}^\ast a_{\we{k}- \we{q}}+c.c.,
\label{H_c}
\end{equation}
\begin{equation}
	H_d=\sum_{\we{q}\we{k}\we{k'}}\left(iD_{\we{q}}\phi(\we{k},\we{q})+
	\frac{1}{2}(\varepsilon_{\we{k}	-\we{q}}-\varepsilon_{\we{k}}+
	\omega_{\we{q}})\phi(\we{k},\we{q})\phi^\ast(\we{k'},\we{q})\right) 
	a_{\we{k}}^\ast a_{\we{k}-\we{q}}a_{\we{k'}-\we{q}}^\ast 
	a_{\we{k'}}+c.c..
\label{H_d}
\end{equation}
$H_b$ is linear in~$b_{\we{q}}^\ast b_{\we{q}}$ but differs from~$H_a$, $H_c$~is part 
of~$H_{int}$ which had been excluded from the~transformation to avoid convergence 
problems, $H_d$~is the~2-electron interaction obtained, whereas~$H_e$ contains terms 
of the~form~$(4,1)$ which represent the~obtained correction to Fr\"ohlich's~Hamiltonian.

The~function~$\phi(\we{k},\we{q})$ remains in the~form~\eqref{phi}, which guarantees 
minimization of the~contribution~$H_c$. 
\section{\label{4}The second transformation}
The~structure of~$H'$ bears similarities to that of~$H_{e-ph}$. Both~$H'$ 
and~$H_{e-ph}$ contain terms describing the~interaction of electrons with phonons: 
the~counterpart of~$H_{int}$ in~$H_{e-ph}$ is~$H_e$ in~$H'$. To estimate the~effect 
of~$H_e$, let us repeatedly apply Fr\"ohlich's~method and perform a~second unitary 
transformation adjusted to eliminate~$H_e$ as far as possible. The form of~$H_e$
suggests to take
\begin{equation*}
	S'=\sum_{\we{u}}\left(\eta_{\we{u}}^\ast b_{\we{u}}^\ast -
	\eta_{\we{u}}b_{\we{u}}+\xi_{\we{u}}^\ast
	b_{\we{u}}^\ast -\xi_{\we{u}}b_{\we{u}}+
	\zeta_{\we{u}}^\ast b_{\we{u}}^\ast -\zeta_{\we{u}}b_{\we{u}}\right),
\end{equation*}
where
\begin{equation}
	\eta_{\we{u}}^\ast=\sum_{\we{l}\we{m}}
	\psi^\ast(\we{l},\we{m},\we{u})a_{\we{l}}^\ast 
	a_{\we{l}}a_{\we{m}-\we{u}}^\ast a_{\we{m}},
\label{eta}
\end{equation}
\begin{equation}
	\xi_{\we{u}}^\ast=\sum_{\we{l}\we{m}\we{t}}
	\chi^\ast(\we{l},\we{m},\we{t},\we{u}) 
	a_{\we{l}-\we{u}}^\ast a_{\we{l}-\we{t}}a_{\we{m}-\we{t}}^\ast a_{\we{m}},
\label{xi}
\end{equation}
\begin{equation}
	\zeta_{\we{u}}^\ast=\sum_{\we{l}\we{m}\we{t}}
	\varphi^\ast(\we{l},\we{m},\we{t},\we{u})
	a_{\we{l}}^\ast a_{\we{l}-\we{t}}a_{\we{m}-\we{t}}^\ast a_{\we{m}+\we{u}}.
\label{zeta}
\end{equation}
The~explicit form of the~functions~$\psi$, $\chi$, $\varphi$ will be found below.

The~interaction~$H_e$ is 3rd~order in~$D_{\we{q}}$, therefore bearing in mind 
Fr\"ohlich's approach, we shall restrict the expansion 
of~$\exp[S'^\ast]H'\exp[S']=\hat{H}$ to terms which are 6th~order in~$D_{\we{q}}$. 
These terms are 
\begin{equation}
	\hat{H}=H'-[S',H']+\frac{1}{2}[S',[S',H_a]] +\ldots \, .
\label{hatH}
\end{equation}
To evaluate the~rhs, one needs the~commutators 
\begin{equation*}
\begin{split}
	\left[S',H_a\right]&=
	\sum_{\we{u}\we{k}\we{l}}b_{\we{u}}^\ast 
	\psi^\ast(\we{l},\we{k},\we{u})
	\left( \varepsilon_{\we{k}}-\varepsilon_{\we{k}-\we{u}}-\omega_{\we{u}}\right) 
	a_{\we{l}}^\ast a_{\we{l}} a_{\we{k}-\we{u}}^\ast a_{\we{k}}\\
	& + \sum_{\we{u}\we{k}\we{l}\we{t}}b_{\we{u}}^\ast
	\chi^\ast(\we{l},\we{k},\we{t},\we{u}) 
	\left(\varepsilon_{\we{k}}-\varepsilon_{\we{k}-\we{t}}+
	\varepsilon_{\we{l}-\we{t}}-\varepsilon_{\we{l}-\we{u}}-\omega_{\we{u}}\right) 
	a_{\we{l}-\we{u}}^\ast a_{\we{l}-\we{t}}a_{\we{k}-\we{t}}^\ast a_{\we{k}}\\
	& + \sum_{\we{u}\we{k}\we{l}\we{t}}b_{\we{u}}^\ast
	\varphi^\ast(\we{l},\we{k},\we{t},\we{u})
	\left( \varepsilon_{\we{k}+\we{u}}-\varepsilon_{\we{k}-\we{t}}+
	\varepsilon_{\we{l}-\we{t}}-\varepsilon_{\we{l}}-\omega_{\we{u}}\right) 
	a_{\we{l}}^\ast a_{\we{l}-\we{t}}a_{\we{k}-\we{t}}^\ast 
	a_{\we{k}+\we{u}}+c.c.,\\
\end{split}
\end{equation*}
\begin{equation*}
	\left[S',H_b\right]=-\sum_{\we{q}\we{k}}b_{\we{q}}^\ast n_{\we{k}} 
	\left(\eta_{\we{q}}^\ast+\xi_{\we{q}}^\ast 
	+ \zeta_{\we{q}}^\ast\right)
	\left( D_{\we{k}\we{q}}+D_{\we{k}\we{q}}^\ast\right)+c.c.,
\end{equation*}
where~$D_{\we{k}\we{q}}$ is defined by~$H_b=\sum_{\we{q}}b_{\we{q}}^\ast b_{\we{q}} 
n_{\we{k}}D_{\we{k}\we{q}}+c.c.$ in Eq.~\eqref{H_b}, 
\begin{equation*}
\begin{split}
	&\left[S',H_c\right]=-\sum_{\we{q}\we{k}}E_{\we{k},\we{q}}
	a_{\we{k}}^\ast a_{\we{k}-\we{q}}\left( \eta_{\we{q}}^\ast+\xi_{\we{q}}^\ast+
	\zeta_{\we{q}}^\ast\right)\\
        &+\sum_{\we{q}\we{k}\we{l}}E_{\we{k},\we{q}}b_{\we{q}}^\ast 
	b_{\we{q}}\Bigl\{ \psi^\ast(\we{l},\we{k},\we{q})
	n_{\we{l}}\left(n_{\we{k}-\we{q}}-n_{\we{k}}\right)+
	\bigl( \psi^\ast(\we{k},\we{l},\we{q})-\psi^\ast(\we{k}-\we{q},\we{l},\we{q}) 
	\bigr) a_{\we{k}}^\ast a_{\we{k}-\we{q}} a_{\we{l}-\we{q}}^\ast 
	a_{\we{l}}\Bigr\}\\
	&+\sum_{\we{q}\we{k}\we{l}\we{t}}b_{\we{q}}^\ast b_{\we{q}}
	\Bigl\{ \bigl( E_{\we{k},\we{q}}\chi^\ast(\we{l},\we{k},\we{t},\we{q})-
	E_{\we{k}-\we{t},\we{q}}\chi^\ast(\we{l},\we{k}-\we{q},\we{t},\we{q}) 
	\bigr) a_{\we{l}-\we{q}}^\ast a_{\we{l}-\we{t}}a_{\we{k}-\we{t}}^\ast 
	a_{\we{k}-\we{q}}\\
	&+ \bigl( E_{\we{k}+\we{q}-\we{t},\we{q}}\chi^\ast(\we{k}+\we{q},\we{l},
	\we{t},\we{q})-E_{\we{k},\we{q}}\chi^\ast(\we{k},\we{l},\we{t},\we{q}) 
	\bigr) a_{\we{k}}^\ast a_{\we{k}-\we{t}}a_{\we{l}-\we{t}}^\ast a_{\we{l}}\\
	&+ \bigl( E_{\we{k}+\we{q}+\we{t},\we{q}}\varphi^\ast(\we{l},\we{k}+\we{t},
	\we{t},\we{q}) - E_{\we{k},\we{q}}\varphi^\ast(\we{l},\we{k}+\we{t}-\we{q},
	\we{t},\we{q})\bigr) a_{\we{l}}^\ast a_{\we{l}-\we{t}}a_{\we{k}}^\ast 
	a_{\we{k}+\we{t}}\\
	&+ \bigl( E_{\we{k},\we{q}}\varphi^\ast(\we{k}+\we{t},\we{l},\we{t},\we{q})-
	E_{\we{k}+\we{t},\we{q}}\varphi^\ast(\we{k}+\we{t}-\we{q},\we{l},\we{t},\we{q}) 
	\bigr) a_{\we{k}+\we{t}}^\ast a_{\we{k}-\we{q}}a_{\we{l}-\we{t}}^\ast 
	a_{\we{l}+\we{q}}\Bigr\}+c.c.,\\
\end{split}
\end{equation*}
where~$E_{\we{k},\we{q}}$ is defined by~$H_c=\sum_{\we{q}\we{k}}b_{\we{q}}a_{\we{k}}^\ast 
a_{\we{k}-\we{q}} E_{\we{k},\we{q}}+c.c.$ in Eq.~\eqref{H_c},
\begin{equation*}
\begin{split}
  	\left[S', H_d \right]& = 
	\sum_{\we{u}\we{q}\we{k}\we{k'}}b_{\we{u}}^\ast
	\left(F_{\we{k},\we{k'},\we{q}} +F_{\we{k'},\we{k},\we{q}}^\ast\right)
	\Bigl\{ a_{\we{k}}^\ast a_{\we{k}-\we{q}} \left[ \eta_{\we{u}}^\ast,a_{\we{k'}-
	\we{q}}^\ast a_{\we{k'}}\right]+
	\left[\eta_{\we{u}}^\ast,a_{\we{k}}^\ast a_{\we{k}-\we{q}}\right]a_{\we{k'}-
	\we{q}}^\ast a_{\we{k'}}\\
	&+a_{\we{k}}^\ast a_{\we{k}-\we{q}}\left[\xi_{\we{u}}^\ast,a_{\we{k'}-
	\we{q}}^\ast a_{\we{k'}}\right]+\left[\xi_{\we{u}}^\ast,a_{\we{k}}^\ast 
	a_{\we{k}-\we{q}}\right]a_{\we{k'} -\we{q}}^\ast a_{\we{k'}}
	+a_{\we{k}}^\ast a_{\we{k}-\we{q}}\left[\zeta_{\we{u}}^\ast,a_{\we{k'}
        -\we{q}}^\ast a_{\we{k'}}\right]+\left[\zeta_{\we{u}}^\ast,a_{\we{k}}^\ast 
	a_{\we{k}-\we{q}} \right]a_{\we{k'}-\we{q}}^\ast a_{\we{k'}}\Bigr\}+c.c.,\\
\end{split}
\end{equation*}
where~$F_{\we{k},\we{k'},\we{q}}$ is defined by~$H_d=\sum_{\we{q}\we{k}\we{k'}}F_{\we{k},
\we{k'},\we{q}}a_{\we{k}}^\ast a_{\we{k}-\we{q}}a_{\we{k'}-\we{q}}^\ast a_{\we{k'}}+
c.c.$ in Eq.~\eqref{H_d},
\begin{equation*}
\begin{split}
	&\left[S',H_e\right]=
	-\sum_{\we{q}\we{k}}A_{\we{k},\we{q}}b_{\we{q}}^\ast b_{\we{q}}
	\Bigl\{ \left[ \eta_{\we{q}},n_{\we{k}}\gamma_{\we{q}}^\ast\right]+
	\left[\xi_{\we{q}},n_{\we{k}}\gamma_{\we{q}}^\ast\right]
	+\left[\zeta_{\we{q}},n_{\we{k}}\gamma_{\we{q}}^\ast\right]\Bigr\}\\ 
	&-\sum_{\we{q}\we{w}\we{k}\we{k'}}B_{\we{k},\we{q},\we{w},\we{k'}} 
	b_{\we{w}}^\ast b_{\we{w}}\Bigl\{ \left[\eta_{\we{w}},a_{\we{k}-\we{w}}^\ast 
	a_{\we{k}-\we{q}}a_{\we{k'}-\we{q}}^\ast a_{\we{k'}}\right]
  	+\left[\xi_{\we{w}},a_{\we{k}-\we{w}}^\ast a_{\we{k}-\we{q}}a_{\we{k'}-
	\we{q}}^\ast a_{\we{k'}}\right] 
  	+\left[\zeta_{\we{w}},a_{\we{k}-\we{w}}^\ast a_{\we{k}-\we{q}}a_{\we{k'}-
	\we{q}}^\ast a_{\we{k'}} \right] \Bigr\} \\
	& - \sum_{\we{q}\we{w}\we{k}\we{k'}}C_{\we{k},\we{q},\we{w},\we{k'}} 
	b_{\we{w}}^\ast b_{\we{w}}\Bigl\{ \left[\eta_{\we{w}},a_{\we{k'}}^\ast 
	a_{\we{k'}-\we{q}}a_{\we{k}- \we{q}}^\ast a_{\we{k}+\we{w}}\right]
   	+ \left[\xi_{\we{w}},a_{\we{k'}}^\ast a_{\we{k'}-\we{q}}a_{\we{k}-\we{q}}^\ast 
	a_{\we{k}+\we{w}}\right] +\left[\zeta_{\we{w}},a_{\we{k'}}^\ast a_{\we{k'}-
	\we{q}}a_{\we{k}-\we{q}}^\ast a_{\we{k}+\we{w}} \right]\Bigr\}\\
   	&-\sum_{\we{q}\we{k}}A_{\we{k},\we{q}}\Bigl\{ \eta_{\we{q}} 
	n_{\we{k}}\gamma_{\we{q}}^\ast+\xi_{\we{q}} n_{\we{k}} \gamma_{\we{q}}^\ast+
	\zeta_{\we{q}}n_{\we{k}}\gamma_{\we{q}}^\ast\Bigr\} \\
	& - \sum_{\we{q}\we{w}\we{k}\we{k'}}B_{\we{k},\we{q},\we{w},\we{k'}}
	\Bigl\{ \eta_{\we{w}}a_{\we{k}-\we{w}}^\ast a_{\we{k}-\we{q}}a_{\we{k'}-
	\we{q}}^\ast a_{\we{k'}}
	+\xi_{\we{w}}a_{\we{k}- \we{w}}^\ast a_{\we{k}-\we{q}}a_{\we{k'}-\we{q}}^\ast 
	a_{\we{k'}}+\zeta_{\we{w}}a_{\we{k}-\we{w}}^\ast a_{\we{k}-\we{q}}a_{\we{k'}-
	\we{q}}^\ast a_{\we{k'}}\Bigr\} \\
	& - \sum_{\we{q}\we{w}\we{k}\we{k'}} 
	C_{\we{k},\we{q},\we{w},\we{k'}}\Bigl\{ \eta_{\we{w}} a_{\we{k'}}^\ast 
	a_{\we{k'}-\we{q}}a_{\we{k}-\we{q}}^\ast a_{\we{k}+\we{w}}
	+ \xi_{\we{w}} a_{\we{k'}}^\ast a_{\we{k'}-\we{q}}a_{\we{k}-\we{q}}^\ast 
	a_{\we{k}+\we{w}}+\zeta_{\we{w}} a_{\we{k'}}^\ast a_{\we{k'}-\we{q}}
	a_{\we{k}-\we{q}}^\ast a_{\we{k}+\we{w}}\Bigr\}+c.c.,\\
\end{split}
\end{equation*}
where~$A_{\we{k},\we{q}}$,~$B_{\we{k},\we{q},\we{w},\we{k'}}$,~$C_{\we{k},\we{q},\we{w},
\we{k'}}$ are given, respectively by Eqs.~\eqref{A}, \eqref{B} i \eqref{C}, 
\begin{equation}
\begin{split}
	&\left[S',\left[S',H_a\right]\right]=
	-\sum_{\we{q}\we{k}\we{k'}}G_{\we{k},\we{k'},\we{q}} 
	b_{\we{q}}^\ast b_{\we{q}} \Bigl\{ \left[\eta_{\we{q}},a_{\we{k'}}^\ast 
	a_{\we{k'}}a_{\we{k}-\we{q}}^\ast a_{\we{k}}\right]+\left[\xi_{\we{q}},
	a_{\we{k'}}^\ast a_{\we{k'}}a_{\we{k}-\we{q}}^\ast a_{\we{k}}\right]
	+\left[\zeta_{\we{q}},a_{\we{k'}}^\ast a_{\we{k'}}a_{\we{k}-\we{q}}^\ast 
	a_{\we{k}} \right] \Bigr\} \\
	& - \sum_{\we{q}\we{k}\we{k'}\we{t}}H_{\we{k},\we{k'},\we{t},\we{q}}
	b_{\we{q}}^\ast b_{\we{q}} \Bigl\{ \left[\eta_{\we{q}},a_{\we{k'}-\we{q}}^\ast 
	a_{\we{k'}- \we{t}}a_{\we{k}-\we{t}}^\ast a_{\we{k}}\right]+
	\left[\xi_{\we{q}},a_{\we{k'}-\we{q}}^\ast a_{\we{k'}-\we{t}}a_{\we{k}-
	\we{t}}^\ast a_{\we{k}}\right]
	+\left[\zeta_{\we{q}},a_{\we{k'}-\we{q}}^\ast a_{\we{k'}-\we{t}} 
	a_{\we{k}-\we{t}}^\ast a_{\we{k}}\right]\Bigr\}\\
	& - \sum_{\we{q}\we{k}\we{k'}\we{t}}I_{\we{k},\we{k'},\we{t},\we{q}} 
	b_{\we{q}}^\ast b_{\we{q}}\Bigl\{ 
	\left[\eta_{\we{q}},a_{\we{k'}}^\ast a_{\we{k'}-\we{t}}a_{\we{k}-\we{t}}^\ast 
	a_{\we{k}+\we{q}}\right]
	+\left[\xi_{\we{q}},a_{\we{k'}}^\ast a_{\we{k'}-\we{t}}a_{\we{k}-\we{t}}^\ast 
	a_{\we{k}+\we{q}} \right]+
	\left[\zeta_{\we{q}},a_{\we{k'}}^\ast a_{\we{k'}-\we{t}}a_{\we{k}-\we{t}}^\ast 
	a_{\we{k}+\we{q}}\right]\Bigr\}\\
	&-\sum_{\we{q}\we{k}\we{k'}}G_{\we{k},\we{k'},\we{q}} \Bigl\{ 
	\eta_{\we{q}} n_{\we{k'}}a_{\we{k}-\we{q}}^\ast a_{\we{k}}+\xi_{\we{q}} 
	n_{\we{k'}}a_{\we{k}-\we{q}}^\ast a_{\we{k}} +\zeta_{\we{q}} n_{\we{k'}}
	a_{\we{k}-\we{q}}^\ast a_{\we{k}}\Bigr\}\\
	&-\sum_{\we{q}\we{k}\we{k'}\we{t}}H_{\we{k},\we{k'},\we{t},\we{q}} \Bigl\{
	\eta_{\we{q}}a_{\we{k'}-\we{q}}^\ast a_{\we{k'}-\we{t}}a_{\we{k}-\we{t}}^\ast 
	a_{\we{k}}+\xi_{\we{q}}a_{\we{k'}-\we{q}}^\ast a_{\we{k'}-\we{t}}a_{\we{k}-
	\we{t}}^\ast a_{\we{k}}+\zeta_{\we{q}}a_{\we{k'}-\we{q}}^\ast a_{\we{k'}-\we{t}}
	a_{\we{k}-\we{t}}^\ast a_{\we{k}}\Bigr\}\\
	&-\sum_{\we{q}\we{k}\we{k'}\we{t}}I_{\we{k},\we{k'},\we{t},\we{q}} \Bigl\{ 
	\eta_{\we{q}} a_{\we{k'}}^\ast a_{\we{k'}-\we{t}}a_{\we{k}-\we{t}}^\ast 
	a_{\we{k}+\we{q}}+\xi_{\we{q}} a_{\we{k'}}^\ast a_{\we{k'}-\we{t}}
	a_{\we{k}-\we{t}}^\ast a_{\we{k}+\we{q}}+\zeta_{\we{q}}
	a_{\we{k'}}^\ast a_{\we{k'}-\we{t}}a_{\we{k}-\we{t}}^\ast a_{\we{k}+\we{q}} 
	\Bigr\}+c.c.,\\
\end{split}
\label{S'S'Ha}
\end{equation}
where
\begin{equation}
 	G_{\we{k},\we{k'},\we{q}}=\psi^\ast(\we{k'},\we{k},\we{q}) 
	\left(\varepsilon_{\we{k}}-\varepsilon_{\we{k}-\we{q}}-
	\omega_{\we{q}}\right),
\label{G}
\end{equation}
\begin{equation}
	H_{\we{k},\we{k'},\we{t},\we{q}}=
	\chi^\ast(\we{k'},\we{k},\we{t},\we{q})
	\left(\varepsilon_{\we{k}}-\varepsilon_{\we{k}-\we{t}}+
	\varepsilon_{\we{k'}-\we{t}}-\varepsilon_{\we{k'}-\we{q}}-\omega_{\we{q}} 
	\right),
\label{H}
\end{equation}
\begin{equation}
	I_{\we{k},\we{k'},\we{t},\we{q}}=
	\varphi^\ast(\we{k'},\we{k},\we{t},\we{q})
	\left(\varepsilon_{\we{k}+\we{q}}-\varepsilon_{\we{k}-\we{t}}+
	\varepsilon_{\we{k'}-\we{t}}-\varepsilon_{\we{k'}}-\omega_{\we{q}} 
	\right).
\label{I}
\end{equation}
The~commutators in the~terms containing phonon operators have not been evaluated, as we 
are interested first of all in expressions containing exclusively electron operators.

Given the~transformed Hamiltonian~$\hat{H}$, we are now in position to minimize 
the~effect of~$H_e$ by imposing, similarly as~Fr\"ohlich, the~condition~$H_e-\left[S',
H_a\right]=0$. This leads to equations for~$\psi$, $\chi$, $\varphi$, which determine these functions uniquely, viz, 
\begin{equation}
	\psi^\ast(\we{k'},\we{k},\we{q})=
	\frac{\phi^\ast(\we{k},\we{q})A_{\we{k'},\we{q}}}{\varepsilon_{\we{k}}
	-\varepsilon_{\we{k}-\we{q}}-\omega_{\we{q}}}.
\label{psi}
\end{equation}
After substituting~$\phi$ and~$A$ given by Eqs.~\eqref{phi},~\eqref{A}, one obtains
\begin{equation}
	\psi^\ast(\we{k'},\we{k},\we{q})=
	\frac{2iD_{\we{q}}^3\left( 1-\Delta(\we{k},\we{q})\right)}{3\left(
  	\varepsilon_{\we{k}-\we{q}}-\varepsilon_{\we{k}}+\omega_{\we{q}}\right)^2}
	\left( \frac{1-\Delta(\we{k'},\we{q})}{\varepsilon_{\we{k'}-\we{q}}-
	\varepsilon_{\we{k'}}+\omega_{\we{q}}}-\frac{1-\Delta(\we{k'}+\we{q},
  	\we{q})}{\varepsilon_{\we{k'}}-\varepsilon_{\we{k'}+\we{q}}+
	\omega_{\we{q}}}\right).
\label{psi_exp}
\end{equation}
Introduction of additional functions in order to preserve convergence is not necessary 
here, as~$\Delta$ already guarantees this property.

As for~$\chi$ and~$\varphi$, additional functions preserving convergence are 
indispensable, viz,
\begin{equation}
	\chi^\ast(\we{k'},\we{k},\we{t},\we{q})=
	\frac{B_{\we{k'},\we{t},\we{q},\we{k}}}{\varepsilon_{\we{k}}-
	\varepsilon_{\we{k}-\we{t}}+\varepsilon_{\we{k'}-\we{t}}-\varepsilon_{\we{k'}-
	\we{q}}-\omega_{\we{q}}} \left( 1-\tilde{\Delta}(\we{k},\we{k'},\we{t},
	\we{q})\right),
\label{chi}
\end{equation}
where
\begin{equation}
	\tilde{\Delta}(\we{k},\we{k'},\we{t},\we{q})=\begin{cases}
	        1 \text{,} & \, \text{if} \quad \left|\varepsilon_{\we{k}}-
		\varepsilon_{\we{k}-\we{t}}+\varepsilon_{\we{k'}-\we{t}}-
		\varepsilon_{\we{k'}-\we{q}}-\omega_{\we{q}}\right| < 
		\tilde{\Gamma}_{\we{q}} \\
       	        0 \text{,} & \, \text{if} \quad \left|\varepsilon_{\we{k}}-
		\varepsilon_{\we{k}-\we{t}}+\varepsilon_{\we{k'}-\we{t}}-
		\varepsilon_{\we{k'}-\we{q}}- \omega_{\we{q}}\right| \ge  
		\tilde{\Gamma}_{\we{q}} \\
	      \end{cases}.
\label{Delta_tilde} 
\end{equation}
Taking into account Eqs.~\eqref{phi},~\eqref{B}, one obtains
\begin{equation}
\begin{split}
	&\chi^\ast(\we{k'},\we{k},\we{t},\we{q})=\frac{iD_{\we{t}}^2 
	D_{\we{q}}}{6\left(\varepsilon_{\we{k}}-\varepsilon_{\we{k}-\we{t}}+
	\varepsilon_{\we{k'}-\we{t}}-\varepsilon_{\we{k'}-\we{q}}-
	\omega_{\we{q}} \right)}
	\left( \frac{\left( 1-\Delta(\we{k},\we{t}) \right)\left(1-\Delta(\we{k'},
	\we{q})\right)}{\left( \varepsilon_{\we{k}-\we{t}}-\varepsilon_{\we{k}}+
	\omega_{\we{t}}\right)\left(
	\varepsilon_{\we{k'}-\we{q}}-\varepsilon_{\we{k'}}+\omega_{\we{q}}\right)}\left( 
	2+\Delta(\we{k'},\we{t})\right)\right.\\
	&-\frac{\left( 1-\Delta(\we{k},\we{t}) \right)\left(1-\Delta(\we{k'}-\we{t},
	\we{q})\right)}{\left(
	\varepsilon_{\we{k}-\we{t}}-\varepsilon_{\we{k}}+\omega_{\we{t}}\right)\left( 	
	\varepsilon_{\we{k'}-
	\we{t}-\we{q}}-\varepsilon_{\we{k'}-\we{t}}+\omega_{\we{q}}\right)}\left( 2+
	\Delta(\we{k'}-\we{q},\we{t})\right)
	+\frac{\left( 1-\Delta(\we{k'},\we{t}) \right)\left(1-\Delta(\we{k'},\we{q}) 
	\right)}{\left(
	\varepsilon_{\we{k'}-\we{t}}-\varepsilon_{\we{k'}}+\omega_{\we{t}}\right)\left( 
	\varepsilon_{\we{k'}
	-\we{q}}-\varepsilon_{\we{k'}}+\omega_{\we{q}}\right)}\left( 4-\Delta(\we{k},
	\we{t})\right)\\
	&\left.-\frac{\left( 1-\Delta(\we{k'}-\we{q},\we{t}) \right)\left(1-
	\Delta(\we{k'}-\we{t},\we{q})
	\right)}{\left( \varepsilon_{\we{k'}-\we{q}-\we{t}}-\varepsilon_{\we{k'}-\we{q}}+
	\omega_{\we{t}}
	\right)\left( \varepsilon_{\we{k'}-\we{t}-\we{q}}-\varepsilon_{\we{k'}-\we{t}}+
	\omega_{\we{q}}
	\right)}\left( 4-\Delta(\we{k},\we{t})\right)\right)\left( 1-
	\tilde{\Delta}(\we{k},\we{k'},\we{t},\we{q})\right),\\
\end{split}
\label{chi_exp}
\end{equation}
\begin{equation}
	\varphi^\ast(\we{k'},\we{k},\we{t},\we{q})=\frac{C_{\we{k},\we{t},\we{q},
	\we{k'}}}{\varepsilon_{\we{k}+\we{q}}-\varepsilon_{\we{k}-\we{t}}+
	\varepsilon_{\we{k'}-\we{t}} -\varepsilon_{\we{k'}}-\omega_{\we{q}}}\left( 1-
	\hat{\Delta}(\we{k},\we{k'},\we{t},\we{q}) \right),
\label{varphi}
\end{equation}
where
\begin{equation}
	\hat{\Delta}(\we{k},\we{k'},\we{t},\we{q})=\begin{cases}
	        1 \text{,} & \, \text{if} \quad \left| \varepsilon_{\we{k}+\we{q}}-
		\varepsilon_{\we{k}-\we{t}}+\varepsilon_{\we{k'}-\we{t}}-
		\varepsilon_{\we{k'}}-\omega_{\we{q}}\right| < \hat{\Gamma}_{\we{q}}\\
	        0 \text{,} & \, \text{if} \quad \left|\varepsilon_{\we{k}+\we{q}}-
		\varepsilon_{\we{k}-\we{t}}+\varepsilon_{\we{k'}-\we{t}}-
		\varepsilon_{\we{k'}}-\omega_{\we{q}}\right| \ge \hat{\Gamma}_{\we{q}}\\
	      \end{cases},
\label{Delta_hat} 
\end{equation}
and substitution of~$\phi$ and~$C_{\we{k},\we{t},\we{q},\we{k'}}$ from 
Eqs.~\eqref{phi},~\eqref{C} yields
\begin{equation}
\begin{split}
	& \varphi^\ast(\we{k'},\we{k},\we{t},\we{q}) = \frac{iD_{\we{t}}^2	
	D_{\we{q}}}{6\left( \varepsilon_{\we{k}+\we{q}}-\varepsilon_{\we{k}-\we{t}}+
	\varepsilon_{\we{k'}-\we{t}}- \varepsilon_{\we{k'}}-\omega_{\we{q}} 
	\right)}
	\left( \frac{\left( 1-\Delta(\we{k'},\we{t})\right)\left(1-\Delta(\we{k}+
	\we{q}-\we{t},\we{q})\right)}{\left(\varepsilon_{\we{k'}-\we{t}}-
	\varepsilon_{\we{k'}}+\omega_{\we{t}}\right)
	\left(\varepsilon_{\we{k}-\we{t}}-\varepsilon_{\we{k}+\we{q}-\we{t}}+
	\omega_{\we{q}}\right)}\left( 2+\Delta(\we{k}+\we{q},\we{t})\right)\right.\\
	&-\frac{\left( 1-\Delta(\we{k'},\we{t})\right)\left(1-\Delta(\we{k}+\we{q},
	\we{q})\right)}{\left( \varepsilon_{\we{k'}-\we{t}}-\varepsilon_{\we{k'}}+
	\omega_{\we{t}}\right)\left(\varepsilon_{\we{k}}-
	\varepsilon_{\we{k}+\we{q}}+\omega_{\we{q}}\right)}\left( 2+\Delta(\we{k},
	\we{t})\right)
	+\frac{\left(1-\Delta(\we{k}+\we{q},\we{t})\right)\left(1-\Delta(\we{k}+\we{q}-
	\we{t}, \we{q})\right)}{\left(\varepsilon_{\we{k}+\we{q}-\we{t}}-
	\varepsilon_{\we{k}+\we{q}}+\omega_{\we{t}}\right)\left(\varepsilon_{\we{k}-
	\we{t}}-\varepsilon_{\we{k}+\we{q}-\we{t}}+\omega_{\we{q}}\right)}\left( 4-
	\Delta(\we{k'},\we{t})\right)\\
	&\left.-\frac{\left( 1-\Delta(\we{k},\we{t})\right)\left(1-\Delta(\we{k}+\we{q},
	\we{q})\right)}{\left(\varepsilon_{\we{k}-\we{t}}-\varepsilon_{\we{k}}+
	\omega_{\we{t}}\right)\left( \varepsilon_{\we{k}}-\varepsilon_{\we{k}+\we{q}}+
	\omega_{\we{q}}\right)}\left( 4-\Delta(\we{k'},\we{t})\right)\right)\left( 1-
	\hat{\Delta}(\we{k},\we{k'},\we{t},\we{q})\right),\\
\end{split}
\label{varphi_exp}
\end{equation}
Having established~$\psi$, $\chi$, $\varphi$, let us average~$\hat{H}$ over the~phonon 
vacuum:
\begin{equation}
\begin{split}
	\hat{H}_{\mathrm{av}}&=\frac{1}{2}\sum_{\we{k}}\varepsilon_{\we{k}}n_{\we{k}}+
	\sum_{\we{q}\we{k}\we{k'}} F_{\we{k},\we{k'},\we{q}}a_{\we{k}}^\ast a_{\we{k}-
	\we{q}}a_{\we{k'}-\we{q}}^\ast a_{\we{k'}}+
	\sum_{\we{q}\we{k}}E_{\we{k},\we{q}}a_{\we{k}}^\ast a_{\we{k}-\we{q}} 
	\left(\eta_{\we{q}}^\ast+ \xi_{\we{q}}^\ast+\zeta_{\we{q}}^\ast\right)\\
	&+\sum_{\we{q}\we{k}\we{k'}}\left( A_{\we{k'},\we{q}}\phi^\ast(\we{k},\we{q})-
	\frac{1}{2} G_{\we{k},\we{k'},\we{q}}\right) 	
	\Bigl\{ \eta_{\we{q}}n_{\we{k'}}a_{\we{k}-\we{q}}^\ast a_{\we{k}}+
	\xi_{\we{q}}n_{\we{k'}}a_{\we{k}-\we{q}}^\ast a_{\we{k}}+\zeta_{\we{q}} 
	n_{\we{k'}}a_{\we{k}- \we{q}}^\ast a_{\we{k}}\Bigr\}\\
	&+\sum_{\we{q}\we{k}\we{k'}\we{t}}\left( B_{\we{k'},\we{t},\we{q},\we{k}}-
	\frac{1}{2}H_{\we{k},\we{k'},\we{t},\we{q}}\right)\Bigl\{\eta_{\we{q}}
	a_{\we{k'}-\we{q}}^\ast a_{\we{k'}-\we{t}}a_{\we{k}
	-\we{t}}^\ast a_{\we{k}}+\xi_{\we{q}}a_{\we{k'}-\we{q}}^\ast a_{\we{k'}-
	\we{t}}a_{\we{k}-\we{t}}^\ast a_{\we{k}}
	+ \zeta_{\we{q}}a_{\we{k'}-\we{q}}^\ast a_{\we{k'}-\we{t}}a_{\we{k}-
	\we{t}}^\ast a_{\we{k}} \Bigr\} \\
	& + \sum_{\we{q}\we{k}\we{k'}\we{t}}\left( 
	C_{\we{k},\we{t},\we{q},\we{k'}}-\frac{1}{2}I_{\we{k},\we{k'},\we{t},
	\we{q}}\right)\Bigl\{\eta_{\we{q}}a_{\we{k'}}^\ast a_{\we{k'}-\we{t}}a_{\we{k}-
	\we{t}}^\ast a_{\we{k}+\we{q}}
	+ \xi_{\we{q}}a_{\we{k'}}^\ast a_{\we{k'}-\we{t}}a_{\we{k}-\we{t}}^\ast
	a_{\we{k}+\we{q}}+\zeta_{\we{q}}a_{\we{k'}}^\ast a_{\we{k'}-\we{t}}a_{\we{k}-
	\we{t}}^\ast a_{\we{k}+\we{q}}\Bigr\}+c.c.,\\
\end{split}
\label{H_av}
\end{equation}

$\hat{H}_{\mathrm{av}}$ contains the~free-electron term, the~2-electron 
Fr\"ohlich~interaction and terms representing 3-electron and 4-electron interactions. 
The~3-electron terms were generated by the~second transformation of~$H_c$. 
The~source of these terms is thus the~non-transformed part of the~original
interaction, discarded by~BCS. If higher-order corrections to BCS~theory are of 
interest, all terms arising from that part, i.e., the~3-electron ones, can be therefore 
neglected. The~same conclusion was drawn above on the grounds of unstability.
\section{\label{5}4-fermion interactions}
The~Hamiltonian~$\hat{H}_{\mathrm{av}}$ contains several terms representing 4-fermion 
interactions. Using Eqs.~\eqref{G}, \eqref{psi}, \eqref{chi}, \eqref{varphi}, one 
finds 
\begin{equation*}
  	A_{\we{k'},\we{q}}\phi^\ast(\we{k},\we{q})-\frac{1}{2}G_{\we{k},\we{k'},\we{q}} 
	=\frac{1}{2}G_{\we{k},\we{k'},\we{q}}=\frac{1}{2}\phi^\ast(\we{k},\we{q}) 
	A_{\we{k'},\we{q}}.
\end{equation*}
Additionally, taking into account Eqs.~\eqref{H},~\eqref{I}, we get
\begin{equation*}
	B_{\we{k'},\we{t},\we{q},\we{k}}-\frac{1}{2}H_{\we{k},\we{k'},\we{t},\we{q}} 
	=\frac{1}{2} H_{\we{k},\we{k'},\we{t},\we{q}}=
	\frac{1}{2}B_{\we{k'},\we{t},\we{q},\we{t}}
	\left( 1-\tilde{\Delta}(\we{k},\we{k'},\we{t},\we{q}) \right),
\end{equation*}
\begin{equation*}
	C_{\we{k},\we{t},\we{q},\we{k'}}-\frac{1}{2}I_{\we{k},\we{k'},\we{t},\we{q}} 
	=\frac{1}{2}I_{\we{k},\we{k'},\we{t},\we{q}}=
	\frac{1}{2}C_{\we{k},\we{t},\we{q},\we{k'}}\left( 1-\hat{\Delta}(\we{k},
	\we{k'},\we{t},\we{q})\right).
\end{equation*}
In terms
of~$G_{\we{k},\we{k'},\we{q}}$,~$H_{\we{k},\we{k'},\we{t},\we{q}}$,~$I_{\we{k},
\we{k'},\we{t},\we{q}}$ the~4-fermion interactions present in~$\hat{H}_{\mathrm{av}}$ 
express as
\begin{equation}
	H_4^1=\sum_{\we{q}\we{k}\we{k'}\we{l}\we{m}}\frac{1}{2}G_{\we{k},\we{k'},	
	\we{q}}\psi(\we{l},\we{m},\we{q})a_{\we{m}}^\ast a_{\we{m}-\we{q}} 
	n_{\we{l}}n_{\we{k'}}a_{\we{k}-\we{q}}^\ast a_{\we{k}}+c.c.,
\label{H41}
\end{equation}
\begin{equation}
	H_4^2=\sum_{\we{q}\we{k}\we{k'}\we{l}\we{m}\we{t}}\frac{1}{2}G_{\we{k},\we{k'},
	\we{q}}\chi(\we{l},\we{m},\we{t},\we{q})a_{\we{m}}^\ast a_{\we{m}-\we{t}} 
	a_{\we{l}-\we{t}}^\ast a_{\we{l}-\we{q}} n_{\we{k'}} a_{\we{k}-\we{q}}^\ast 
	a_{\we{k}}+c.c.,
\label{H42}
\end{equation}
\begin{equation}
	H_4^3=\sum_{\we{q}\we{k}\we{k'}\we{l}\we{m}\we{t}}\frac{1}{2}G_{\we{k},\we{k'},
	\we{q}}\varphi(\we{l},\we{m},\we{t},\we{q})a_{\we{m}+\we{q}}^\ast a_{\we{m}-
	\we{t}}a_{\we{l}-\we{t}}^\ast a_{\we{l}}n_{\we{k'}}a_{\we{k}-\we{q}}^\ast 
	a_{\we{k}}+c.c.,
\label{H43}
\end{equation}
\begin{equation}
	H_4^4=\sum_{\we{q}\we{k}\we{k'}\we{l}\we{m}\we{t}}\frac{1}{2}H_{\we{k},\we{k'},
	\we{t},\we{q}}\psi(\we{l},\we{m},\we{q})a_{\we{m}}^\ast a_{\we{m}-\we{q}} 
	n_{\we{l}}a_{\we{k'}-\we{q}}^\ast a_{\we{k'}-\we{t}}a_{\we{k}-\we{t}}^\ast 
	a_{\we{k}}+c.c.,
\label{H44}
\end{equation}
\begin{equation}
	H_4^5=\sum_{\we{q}\we{k}\we{k'}\we{l}\we{m}\we{t}\we{w}}\frac{1}{2}H_{\we{k},
	\we{k'},\we{t},\we{q}} \chi(\we{l},\we{m},\we{w},\we{q})a_{\we{m}}^\ast 
	a_{\we{m}-\we{w}}a_{\we{l}-\we{w}}^\ast a_{\we{l}-\we{q}} a_{\we{k'}-\we{q}}^\ast 
	a_{\we{k'}-\we{t}}a_{\we{k}-\we{t}}^\ast a_{\we{k}}+c.c.,
\label{H45}
\end{equation}
\begin{equation}
	H_4^6=\sum_{\we{q}\we{k}\we{k'}\we{l}\we{m}\we{t}\we{w}}\frac{1}{2}H_{\we{k},
	\we{k'},\we{t},\we{q}} \varphi(\we{l},\we{m},\we{w},\we{q})a_{\we{m}+\we{q}}^\ast 
	a_{\we{m}-\we{w}}a_{\we{l}-\we{w}}^\ast a_{\we{l}} a_{\we{k'}-\we{q}}^\ast 	
	a_{\we{k'}-\we{t}}a_{\we{k}-\we{t}}^\ast a_{\we{k}}+c.c.,
\label{H46}
\end{equation}
\begin{equation}
	H_4^7=\sum_{\we{q}\we{k}\we{k'}\we{l}\we{m}\we{t}}\frac{1}{2}I_{\we{k},\we{k'},
	\we{t},\we{q}} \psi(\we{l},\we{m},\we{q})a_{\we{m}}^\ast a_{\we{m}-\we{q}} 
	n_{\we{l}}a_{\we{k'}}^\ast a_{\we{k'}-\we{t}} a_{\we{k}-\we{t}}^\ast a_{\we{k}+
	\we{q}}+c.c.,
\label{H47}
\end{equation}
\begin{equation}
	H_4^8=\sum_{\we{q}\we{k}\we{k'}\we{l}\we{m}\we{t}\we{w}}\frac{1}{2}I_{\we{k},
	\we{k'},\we{t},\we{q}} \chi(\we{l},\we{m},\we{w},\we{q})a_{\we{m}}^\ast 
	a_{\we{m}-\we{w}}a_{\we{l}-\we{w}}^\ast a_{\we{l}-\we{q}} 
	a_{\we{k'}}^\ast a_{\we{k'}-\we{t}}a_{\we{k}-\we{t}}^\ast a_{\we{k}+\we{q}}+c.c.,
\label{H48}
\end{equation}
\begin{equation}
	H_4^9=\sum_{\we{q}\we{k}\we{k'}\we{l}\we{m}\we{t}\we{w}}\frac{1}{2}I_{\we{k},
	\we{k'},\we{t},\we{q}} \varphi(\we{l},\we{m},\we{w},\we{q})a_{\we{m}+\we{q}}^\ast 
	a_{\we{m}-\we{w}}a_{\we{l}-\we{w}}^\ast a_{\we{l}} a_{\we{k'}}^\ast 
	a_{\we{k'}-\we{t}}a_{\we{k}-\we{t}}^\ast a_{\we{k}+\we{q}}+c.c..
\label{H49}
\end{equation}
	 
Five of these interactions, viz,~$H_4^1$, $H_4^2$, $H_4^3$, $H_4^4$, $H_4^7$, contain 
the~operator~$n_{\we{k}}$, therefore they are not reducible to the 4-fermion MT 
potential~\eqref{V}. This potential is particularly interesting not only for its 
possible relevance to the~physics of superconductors, but also because 
the~thermodynamics of the~Hamiltonian~$H_{BCS}+W+V_{MT}$ is exactly solvable 
(Brankov~et~al.\cite{brankov}).

Let us consider the~8-fold product of operators in~$H_4^5$ with explicit spin indices:
\begin{equation*}
	a_{\we{m}\sigma''}^\ast a_{\we{m}-\we{w}\sigma''}a_{\we{l}-\we{w}\sigma'''}^\ast 
	a_{\we{l}-\we{q}\sigma'''} a_{\we{k'}-\we{q}\sigma}^\ast 
	a_{\we{k'}-\we{t}\sigma}a_{\we{k}-\we{t}\sigma'}^\ast 
	a_{\we{k}\sigma'}.
\end{equation*}
With respect to 1-fermion momenta these are nine possible reductions to 
the~form~\eqref{V}, and each of them allows two or four possibilities related to spin 
reduction. The~values assumed by each momentum index are collected in 
Table~\ref{tab1}. In four cases~$\we{q}=0$ and the~coupling vanishes, 
since~$\chi(\we{l},\we{m},\we{w},0)=0$, for all~$\we{l}$, $\we{m}$, $\we{w}$. 
As a~consequence, $H_4^5$~assumes the~reduced form 
\begin{equation}
\begin{split}
        H^5_{4(red)}&=\sum_{\we{k}\ne\we{m}}
        \Bigl\{ 2H_{\we{k},2\we{k}+\we{m},\we{k}+\we{m},2\we{k}+2\we{m}}
        \, \chi(2\we{m}+\we{k},\we{m},\we{k}+\we{m},2\we{k}+2\we{m})\\
	& - H_{\we{k},\we{m},\we{k}+\we{m},2\we{m}}\bigl(\chi(2\we{m}+\we{k},\we{m},
	\we{k}+\we{m},2\we{m})+
	\chi(2\we{m}-\we{k},\we{m},\we{m}-\we{k},2\we{m})\bigr)\\
	& - \chi(\we{k},\we{m},\we{k}+\we{m},2\we{k})\left( H_{\we{k},2\we{k}+\we{m},
	\we{k}+\we{m},2\we{k}} + H_{\we{k},2\we{k}-\we{m},\we{k}-\we{m},2\we{k}} 
	\right)\Bigr\}
   	a_{\we{m}-}^\ast  a_{\we{m}+}^\ast a_{-\we{m}-}^\ast  a_{-\we{m}+}^\ast 
	a_{-\we{k}+}a_{-\we{k}-} a_{\we{k}+}a_{\we{k}-}+ c.c..\\ 
\end{split}
\label{red1}
\end{equation}
Terms with~$\we{k}=\we{m}$ have been excluded, similarly as in~$V_{BCS}$. Their 
contribution is accounted for by a~shift of 1-fermion energies. The~coefficients on 
the~rhs result after performing summation over spins.

\begin{table}
\begin{tabular}{|c|c|c|c|c|c|l|}
	\hline
	$H_4^5$&$\we{l}$&$\we{k'}$&$\we{q}$&$\we{t}$&$\we{w}$&spins\\\hline
	1&$2\we{m}+\we{k}$&$2\we{k}+\we{m}$&$2\we{k}+2\we{m}$&$\we{k}+\we{m}$&$\we{k}+
	\we{m}$&$\sigma''=-\sigma'''\text{,}\sigma=-\sigma'$\\\hline
	2&$2\we{m}+\we{k}$&$\we{m}$&$2\we{m}$&$\we{k}+\we{m}$&$\we{k}+\we{m}$&$\sigma''=-
	\sigma'''=-\sigma=\sigma'$\\\hline
	3&$2\we{m}-\we{k}$&$\we{m}$&$2\we{m}$&$\we{k}+\we{m}$&$\we{m}-\we{k}$&$\sigma''=-
	\sigma'''=\sigma=-\sigma'$\\\hline
	4&$\we{k}$&$2\we{k}+\we{m}$&$2\we{k}$&$\we{k}+\we{m}$&$\we{k}+\we{m}$&$\sigma''=-
	\sigma=-\sigma'''=\sigma'$\\\hline
	5&$\we{k}$&$\we{m}$&$0$&$\we{k}+\we{m}$&$\we{k}+\we{m}$&$\sigma''=-\sigma\text{,
	}\sigma'''=-\sigma'$\\\hline
	6&$-\we{k}$&$\we{m}$&$0$&$\we{k}+\we{m}$&$\we{m}-\we{k}$&$\sigma''=-\sigma= 
	\sigma'''=-\sigma'$\\\hline
	7&$\we{k}$&$2\we{k}-\we{m}$&$2\we{k}$&$\we{k}-\we{m}$&$\we{k}+\we{m}$&$\sigma''=-
	\sigma'=\sigma=-\sigma'''$\\\hline
	8&$\we{k}$&$-\we{m}$&$0$&$\we{k}-\we{m}$&$\we{k}+\we{m}$&$\sigma''=-\sigma'=-
	\sigma=\sigma'''$\\\hline
	9&$-\we{k}$&$-\we{m}$&$0$&$\we{k}-\we{m}$&$\we{m}-\we{k}$&$\sigma''=-\sigma' 
	\text{,}\sigma'''=-\sigma$\\\hline
\end{tabular}
\caption{Reductions of $H_4^5$}
\label{tab1}
\end{table}

This procedure has been also applied to~$H_4^6$, $H_4^8$ $H_4^9$. The~corresponding 
values of 1-fermion momenta are given in Tables~\ref{tab2}, \ref{tab3}, \ref{tab4}. 
The~additional remark~$\we{k}\to\we{k}-2\we{m}$ (or similar), indicates that 
a~translation of one momentum index is necessary after reduction.

\begin{table}
\begin{tabular}{|c|c|c|c|c|c|l|c|}
	\hline
	$H_4^6$&$\we{l}$&$\we{k'}$&$\we{q}$&$\we{t}$&$\we{w}$&spins&\\\hline
	1&$-\we{k}$&$-\we{m}$&$-2\we{k}-2\we{m}$&-$\we{k}-\we{m}$&$\we{k}+
	\we{m}$&$\sigma''=-\sigma'''\text{,}\sigma=-\sigma'$&$\we{m}\to\we{m}-
	2\we{k}$\\\hline
	2&$\we{k}$&$-\we{m}$&$-2\we{m}$&$\we{k}-\we{m}$&$\we{k}+\we{m}$&$\sigma''=-
	\sigma'''=-\sigma=\sigma'$&\\\hline
	3&$-\we{k}$&$-\we{m}$&$-2\we{m}$&$\we{k}-\we{m}$&$\we{m}-\we{k}$&$\sigma''=-
	\sigma'''=\sigma=-\sigma'$&\\\hline
	4&$-\we{k}$&$4\we{k}+\we{m}$&$2\we{k}$&$3\we{k}+\we{m}$&$\we{k}+\we{m}$&$ 
	\sigma''=-\sigma=-\sigma'''=\sigma'$&$\we{m}\to\we{m}-2\we{k}$\\\hline
	5&$\we{k}$&$\we{m}$&$0$&$\we{k}+\we{m}$&$\we{k}+\we{m}$&$\sigma''=-\sigma\text{,
	}\sigma'''=-\sigma'$&\\\hline
	6&$-\we{k}$&$\we{m}$&$0$&$\we{k}+\we{m}$&$\we{m}-\we{k}$&$\sigma''=-\sigma= 
	\sigma'''=-\sigma'$&\\\hline
	7&$-\we{k}$&$-\we{m}$&$2\we{k}$&$-\we{k}-\we{m}$&$\we{k}+\we{m}$&$\sigma''=-
	\sigma'=\sigma=-\sigma'''$&$\we{m}\to\we{m}-2\we{k}$\\\hline
	8&$\we{k}$&$-\we{m}$&$0$&$\we{k}-\we{m}$&$\we{k}+\we{m}$&$\sigma''=-\sigma'=-
	\sigma=\sigma'''$&\\\hline
	9&$-\we{k}$&$-\we{m}$&$0$&$\we{k}-\we{m}$&$\we{m}-\we{k}$&$\sigma''=-\sigma' 
	\text{,}\sigma'''=-\sigma$&\\\hline
\end{tabular}
\caption{Reductions of $H_4^6$}
\label{tab2}
\end{table}

\begin{table}
\begin{tabular}{|c|c|c|c|c|c|l|c|}
	\hline
	$H_4^8$&$\we{l}$&$\we{k'}$&$\we{q}$&$\we{t}$&$\we{w}$&spins&\\\hline
	1&$-\we{k}$&$-\we{m}$&$-2\we{k}-2\we{m}$&$\we{k}+\we{m}$&$-\we{k}-\we{m}$&$ 
	\sigma''=-\sigma'''\text{,}\sigma=-\sigma'$&$\we{k}\to\we{k}-2\we{m}$\\\hline
	2&$\we{k}+4\we{m}$&$-\we{m}$&$2\we{m}$&$\we{k}+\we{m}$&$\we{k}+3\we{m}$&$ 
	\sigma''=-\sigma'''=-\sigma=\sigma'$&$\we{k}\to\we{k}-2\we{m}$\\\hline
	3&$-\we{k}$&$-\we{m}$&$2\we{m}$&$\we{k}+\we{m}$&$-\we{m}-\we{k}$&$\sigma''=-
	\sigma'''=\sigma=-\sigma'$&$\we{k}\to\we{k}-2\we{m}$\\\hline
	4&$-\we{k}$&$\we{m}$&$-2\we{k}$&$\we{k}+\we{m}$&$-\we{k}+\we{m}$&$\sigma''=-
	\sigma=\sigma'=-\sigma'''$&\\\hline
	5&$\we{k}$&$\we{m}$&$0$&$\we{k}+\we{m}$&$\we{k}+\we{m}$&$\sigma''=-\sigma\text{,
	}\sigma'''=-\sigma'$&\\\hline
	6&$-\we{k}$&$\we{m}$&$0$&$\we{k}+\we{m}$&$\we{m}-\we{k}$&$\sigma''=-\sigma=-
	\sigma'=\sigma'''$&\\\hline
	7&$-\we{k}$&$-\we{m}$&$-2\we{k}$&$\we{k}-\we{m}$&$-\we{k}+\we{m}$&$\sigma''=-
	\sigma'=\sigma=-\sigma'''$&\\\hline
	8&$\we{k}$&$-\we{m}$&$0$&$\we{k}-\we{m}$&$\we{k}+\we{m}$&$\sigma''=-\sigma'=-
	\sigma=\sigma'''$&\\\hline
	9&$-\we{k}$&$-\we{m}$&$0$&$\we{k}-\we{m}$&$\we{m}-\we{k}$&$\sigma''=-\sigma' 
	\text{,}\sigma'''=-\sigma$&\\\hline
\end{tabular}
\caption{Reductions of $H_4^8$}
\label{tab3}
\end{table}

\begin{table}
\begin{tabular}{|c|c|c|c|c|c|l|l|}
\hline
	$H_4^9$&$\we{l}$&$\we{k'}$&$\we{q}$&$\we{t}$&$\we{w}$&spins &\\\hline
	1&$-\frac{1}{3}\we{k}+\frac{2}{3}\we{m}$&$\frac{2}{3}\we{k}-\frac{1}{3}\we{m}$&$-
	\frac{2}{3}\we{k}-\frac{2}{3}\we{m}$&$\frac{1}{3}\we{k}+\frac{1}{3} 
	\we{m}$&$\frac{1}{3}\we{k}+\frac{1}{3}\we{m}$&$\sigma''=-\sigma'''\text{,}\sigma
	=-\sigma'$&$\we{k}\to3\we{k}+2\we{m}\text{,}\we{m}\to\we{m}-2\we{k}$\\\hline
	2&$\we{k}-2\we{m}$&$\we{m}$&$-2\we{m}$&$\we{k}-\we{m}$&$\we{k}-\we{m}$&$\sigma''=-
	\sigma'''=-\sigma=\sigma'$&$\we{k}\to\we{k}+2\we{m}$\\\hline
	3&$2\we{m}-\we{k}$&$\we{m}$&$-2\we{m}$&$\we{k}-\we{m}$&$3\we{m}-\we{k}$&$\sigma'' 
	=-\sigma'''=\sigma=-\sigma'$&$\we{k}\to\we{k}+2\we{m}$\\\hline
	4&$\we{k}$&$-2\we{k}+\we{m}$&$-2\we{k}$&$-\we{k}+\we{m}$&$-\we{k}+\we{m}$&$ 
	\sigma''=-\sigma=\sigma'=-\sigma'''$&$\we{m}\to\we{m}+2\we{k}$\\\hline
	5&$\we{k}$&$\we{m}$&$0$&$\we{k}+\we{m}$&$\we{k}+\we{m}$&$\sigma''=-\sigma\text{,
	}\sigma'''=-\sigma'$&\\\hline
	6&$-\we{k}$&$\we{m}$&$0$&$\we{k}+\we{m}$&$\we{m}-\we{k}$&$\sigma''=-\sigma=-
	\sigma'=\sigma'''$&\\\hline
	7&$\we{k}$&$2\we{k}-\we{m}$&$-2\we{k}$&$3\we{k}-\we{m}$&$-\we{k}+\we{m}$&$ 
	\sigma''=-\sigma'=\sigma=-\sigma'''$&$\we{m}\to\we{m}+2\we{k}$\\\hline
	8&$\we{k}$&$-\we{m}$&$0$&$\we{k}-\we{m}$&$\we{k}+\we{m}$&$\sigma''=-\sigma'=-
	\sigma=\sigma'''$&\\\hline
	9&$-\we{k}$&$-\we{m}$&$0$&$\we{k}-\we{m}$&$\we{m}-\we{k}$&$\sigma''=-\sigma' 
	\text{,}\sigma'''=-\sigma$&\\\hline
\end{tabular}
\caption{Reductions of $H_4^9$}
\label{tab4}
\end{table}

After reduction, $H_4^6$, $H_4^8$, $H_4^9$~take the~following forms:
\begin{equation}
\begin{split}
	& H^6_{4(red)}=\sum_{\we{k}\ne\we{m}} \Bigl\{ 2H_{\we{k},2\we{k}+\we{m},\we{k}+
	\we{m},2\we{k}+2\we{m}} \, \varphi(-\we{k},-2\we{k}-\we{m},-\we{k}-\we{m},2\we{k}
	+2\we{m})\\
	&-H_{\we{k},\we{m},\we{k}+\we{m},2\we{m}}\bigl( \varphi(\we{k},-\we{m},\we{k}-
	\we{m},2\we{m})+ \varphi(-\we{k},-\we{m},-\we{m}-\we{k},2\we{m})\bigr)\\
	&-\varphi(-\we{k},\we{m}-2\we{k},-\we{k}+\we{m},2\we{k})\bigl( H_{\we{k},2\we{k}+
	\we{m},\we{k}+\we{m},2\we{k}}+H_{\we{k},2\we{k}-\we{m},\we{k}-\we{m},2\we{k}} 
	\bigr) \Bigr\}
	a_{\we{m}-}^\ast  a_{\we{m}+}^\ast  a_{-\we{m}-}^\ast  a_{-\we{m}+}^\ast 
	a_{-\we{k}+}a_{-\we{k}-} a_{\we{k}+} a_{\we{k}-}+c.c.,\\
\end{split}
\label{red2}
\end{equation}
\begin{equation}
\begin{split}
	H^8_{4(red)}&=\sum_{\we{k}\ne\we{m}} \Bigl\{ 2I_{-\we{k}-2\we{m},-\we{m},-\we{k}-
	\we{m},2\we{k}+ 2\we{m}} \, \chi(\we{k}+2\we{m},\we{m},\we{k}+\we{m},2\we{k}+
	2\we{m})\\
	&-I_{\we{k}-2\we{m},-\we{m},\we{k}-\we{m},2\we{m}}\bigl(\chi(\we{k}+2\we{m},
	\we{m},\we{k}+\we{m},2\we{m})+\chi(-\we{k}+2\we{m},\we{m},\we{m}-\we{k},2\we{m}) 
	\bigr)\\
	&-\chi(\we{k},\we{m},\we{k}+\we{m},2\we{k})\bigl( I_{-\we{k},\we{m},-\we{k}+
	\we{m},2\we{k}}+ I_{-\we{k},-\we{m},-\we{k}-\we{m},2\we{k}}\bigr)\Bigr\}
	a_{\we{m}-}^\ast  a_{\we{m}+}^\ast  a_{-\we{m}-}^\ast  a_{-\we{m}+}^\ast 
	a_{-\we{k}+}a_{-\we{k}-} a_{\we{k}+} a_{\we{k}-} + c.c.,\\
\end{split}
\label{red3}
\end{equation}
\begin{equation}
\begin{split}
	& H^9_{4(red)}=\sum_{\we{k}\ne\we{m}} \Bigl\{ 2I_{-\we{k}-2\we{m},-\we{m},-\we{k}-
	\we{m},2\we{k}+	2\we{m}} \, \varphi(-\we{k},-2\we{k}-\we{m},-\we{k}-\we{m},
	2\we{k}+2\we{m})\\
	&-I_{\we{k}-2\we{m},-\we{m},\we{k}-\we{m},2\we{m}}\bigl( \varphi(\we{k},-\we{m},
	\we{k}-\we{m},2\we{m}) + \varphi(-\we{k},-\we{m},-\we{k}-\we{m},2\we{m})\bigr)\\
	&-\varphi(-\we{k},-2\we{k}+\we{m},-\we{k}+\we{m},2\we{k})\bigl( I_{-\we{k},
	\we{m},-\we{k}+\we{m},2\we{k}}+I_{-\we{k},-\we{m},-\we{k}-\we{m},2\we{k}} 
	\bigr)\Bigr\}
	\times a_{\we{m}-}^\ast  a_{\we{m}+}^\ast  a_{-\we{m}-}^\ast  a_{-\we{m}+}^\ast 
	a_{-\we{k}+} a_{-\we{k}-} a_{\we{k}+} a_{\we{k}-} + c.c..\\
\end{split}
\label{red4}
\end{equation}
Collecting all terms in Eqs.~\eqref{red1}--\eqref{red4}, one obtains a~$V_{MT}$ 
interaction as in Eq.~\eqref{V} with
\begin{equation}
        g_{\we{m}\we{k}}=-4\omega_{2\we{k}+2\we{m}} 
	\Lambda_{\we{k}\we{m}}\Lambda_{\we{m}\we{k}}^\ast+
	2\Theta_{\we{k}\we{m}}+2\Theta_{\we{m}\we{k}}^\ast,
\label{gmk}
\end{equation}
where
\begin{equation}
        \Lambda_{\we{k}\we{m}}=\chi^\ast(2\we{k}+\we{m},\we{k},\we{k}+\we{m},2\we{k}+
	2\we{m}) + \varphi^\ast(-\we{m},-\we{k}-2\we{m},-\we{k}-\we{m},2\we{k}+2\we{m}),
\label{Lambda}
\end{equation}
\begin{equation}
\begin{split}
	& \Theta_{\we{k}\we{m}} =\omega_{\we{m}}\bigl( \chi^\ast(\we{m},\we{k},\we{k}+
	\we{m},2\we{m})+\varphi^\ast(-\we{m},\we{k}-2\we{m},\we{k}-\we{m},2\we{m}) 
	\bigr) \\
       &\times \Bigl( \chi(2\we{m}+\we{k},\we{m},\we{k}+\we{m},2\we{m})+\chi(2\we{m}-
       \we{k},\we{m},\we{m}-\we{k},2\we{m})
       +\varphi(\we{k},-\we{m},\we{k}-\we{m},2\we{m})+\varphi(-\we{k},-\we{m},-\we{k}-
       \we{m},2\we{m})\Bigr).\\
\end{split}
\label{Theta}
\end{equation}
\section{4-fermion interaction coupling}
The~most important question about the~4-fermion interaction is positive- or
negative-valuedness of~$g_{\we{m}\we{k}}$. In particular, it would be of interest to
determine the~domain of~$g_{\we{m}\we{k}}$ in momentum space where it is attractive.
Unfortunately, the~form of~$g_{\we{m}\we{k}}$ given by Eq.~\eqref{gmk} is extremely 
complicated, so this problem cannot be resolved in general. 

First, let us specify the~quantities occurring in Eq.~\eqref{gmk}. We assume that
$\varepsilon_{\we{k}}=a k^2$, $\omega_{\we{k}} = b k$, $\tilde{\Gamma}_{\we{k}} = c k$,
$\Gamma_{\we{k}}= d k$ and $\hat{\Gamma}_{\we{k}} = e k$, where $a$, $b$, $c$, $d$,
$e$ are real, positive constants.

Following Davydov\cite{davydov}, we apply some approximations in order to 
estimate~$g_{\we{m}\we{k}}$. We assume that the~most significant contribution to 
the~4-fermion~interaction (analogously as in the~BCS~theory) comes from 1-electron
momenta $\we{m}$, $\we{k}$ which satisfy the condition $m \approx k$.

The~case of exact equality is, of course, excluded (in accord with the~restriction 
on summation in Eqs.~\eqref{red1} -- \eqref{red4}), so we use $g_{\we{k}\we{k}}$ only 
as an~abbreviation for~$g_{\we{m}\we{k}}$ under our approximation $m \approx k$.

Thus, the~interaction coupling~$g_{\we{k}\we{k}}$ is given by
\begin{equation}
	g_{\we{k}\we{k}}=-4\omega_{4\we{k}}\Lambda_{\we{k}\we{k}}   
    	\Lambda_{\we{k}\we{k}}^\ast + 2\Theta_{\we{k}\we{k}}+
    	2\Theta_{\we{k}\we{k}}^\ast.
\label{gkk}
\end{equation}
We take into account only vectors with almost compatible directions, otherwise
$g_{\we{k}\we{k}}$~vanishes.

We can thus rewrite Eq.~\eqref{gkk} in the~form:
\begin{equation}
	g_{\we{k}\we{k}}=-16\omega_{\we{k}}|\Lambda_{\we{k}\we{k}}|^2   
  	+ 4 \Re \Theta_{\we{k}\we{k}},
\label{gkk1}
\end{equation}
where the~first term is always nonpositive and $\Re$ means real part. The~second 
term requires detailed analysis. Detailed calculations are performed in Appendix~\ref{c}. 
We have  found~$g_{\we{k}\we{k}}$ in all cases for different values of 
the~constants~$a$, $b$, $c$, $d$, $e$. Most interesting is the case $b>c$, 
$b>e$ and $b > d$, because the~essential part of initial Hamiltonian is transformed
under these conditions. Under a~further approximation, we find the form 
of~$g_{\we{k}\we{k}}$ for~$k \approx k_F$, where $k_F$~is Fermi~momentum.
Additionally we put $c=d=e$ and $\varepsilon_{\we{k}_F} \gg \omega_{\we{k}_F}$
(this is at least true for metals). Under these assumptions 
\begin{equation*}
	\Theta_{\we{k}_F \we{k}_F} \approx 
	\frac{-D_{\we{k}_F}^6}{4\omega_{\we{k}_F}^5}, \qquad 
	\Lambda_{\we{k}_F \we{k}_F}\approx 
	\frac{iD_{\we{k}_F}^3}{12\omega_{\we{k}_F}^2 \varepsilon_{\we{k}_F}},
\end{equation*}
which implies
\begin{equation*}
	g_{\we{k}_F \we{k}_F} \approx \frac{-D_{\we{k}_F}^6}{\omega_{\we{k}_F}^5}
	\label{g_kF} < 0. 
\end{equation*}

Now we can compare the~magnitudes of the~coupling constants of
4-fermion interaction and 2-fermion interaction (under the~same assumptions).
It is shown in Appendix~\ref{a} that $G_{\we{k}_F\we{k}_F}=
-D_{\we{k}_F}^2 / \omega_{\we{k}_F}$. Thus the~interaction
coupling for 4- and 2-fermion (BCS) interactions fulfils the~relation 
$g_{\we{k}_F \we{k}_F}=G_{\we{k}_F\we{k}_F}^3 / \omega_{\we{k}_F}^2$. As a~consequence, 
for a~strong pairing, there is a~significant contribution from 4-fermion interactions. 

Moreover, the~4-fermion interaction coupling~$g_{\we{k}\we{k}}$ is
also negative in most cases without imposing the~approximations~$c=d=e$ 
and~$k\approx k_F$. This is shown in Appendix~\ref{c}.
\section{\label{6}4-fermion interactions in Fr\"ohlich's expansion}
As demonstrated in Section~\ref{3}, 4-fermion interaction terms appear in sixth order 
of the~expansion of Fr\"ohlich's~original transformation. These terms are derived in 
Appendix~\ref{b}. The~various resulting 4-fermion terms considerably outnumber those 
obtained by applying a~second~transformation in Section~\ref{5}. Again a~reduction 
procedure to~$V_{MT}$ is possible for all 4-fermion expressions~\eqref{fr_6ord}, which 
do~not contain the~particle number operator and possess three phonon indices. However,
the~resulting couplings are also complicated functions and, therefore, will not be 
examined in detail.

The~disadvantage of this method is appearance of 3-fermion interactions, which in 
the~double transformation method gave a~small contribution. Here additional arguments 
must be used to discard these terms. 
\section{\label{7}Concluding remarks}
We have extended Fr\"ohlich's~transformation of~$H_{\mathrm{e-ph}}$ to higher order 
terms. This has been done by performing a~second transform of the~first terms discarded 
by~Fr\"ohlich. The~resulting interactions are of 3- and 4-fermion type. The~3-fermion 
terms can be expected to be inessential in superconductivity because of their instability 
under time inversion. The~4-fermion terms are, in general, reducible 
to~$V_{\mathrm{MT}}$, a~BCS-type 4-fermion interaction. The~resulting 4-fermion coupling 
is extremely complicated, but under reasonable approximations it~is negative-valued. 
Moreover, for 1-fermion momenta in the~neighbourhood of Fermi~momentum, it~has the~simple 
form~$-D_{\we{k}_F}^6 / \omega_{\we{k}_F}^5 = G_{\we{k}_F\we{k}_F}^3 /
\omega_{\we{k}_F}^2$, where $G_{\we{k}_F\we{k}_F}$ is the~2-fermion
coupling. This fact implies that 4-fermion interactions are significant
for systems with strong pairing and allows to estimate (relative to 2-fermion coupling 
and phonon energy at Fermi~momentum) their magnitude.

As in BCS~theory, where the~relation between the~gap parameter~$\Delta$ and coupling 
constant~$G$ of the BCS~interaction allows to estimate the~magnitude of~$G$, it can be 
expected that detailed thermodynamics of~$H_{MT}$ will provide a~relation 
between~$|g_{\we{m}\we{k}}|$ and other parameters of the~theory, thereby allowing to
estimate the~magnitude of~$g_{\we{m}\we{k}}$. Another emerging question is Cooper's 
problem\cite{cooper} for a~bound quadruple in the~presence of~$V_{\mathrm{MT}}$
or~$V_{BCS}+V_{MT}$. Kamei and~Miyake deal with this question in a~recent work\cite{kamei}.

The~double transformation has unveiled the~structure of 3-, 4-, 5-fermion interactions. 
Since they are proportional to 4th, 6th, 8th~power of~$D_{\we{w}}$, they are relatively 
weak, so inclusion of these terms, apart from 4-fermion ones, appears unjustified at 
present, although Schneider~et~al.\cite{schn:2} suggest existence of both quadruples and 
sextets in some HTSC.

The~higher-order expansion terms of the~transformed~$H_{\mathrm{e-ph}}$, including 
quadruple, sextet etc.\  interactions can be expected to reveal themselves in materials 
with extremely strong pairing correlations between spin $1/2$ fermions, since 
the~presence of strongly correlated pairs implies at least some kind of weak interaction 
between them. 

On the other hand, our extension of Fr\"ohlich's~transformation shows that if phonon 
mediation exists in a~superconductor, the~4-fermion and 6-fermion interactions are always 
present as supplementary to the~BCS~one. Unfortunately, the~Hamiltonian~$H_{\mathrm{BCS}}+
V_{\mathrm{MT}}$, although exactly solvable, leads to intricate mean-field equations. 

Our results obtained can be generalized in many respects. First of all, the~process of 
averaging over phonon vacuum could be replaced by averaging over the~phonon equilibrium 
state, which could be justified at higher temperatures. This would already lead to 
an~additional 1-fermion term in Fr\"ohlich's~Hamiltonian and modification of 2-fermion 
and 3-fermion terms in our method. Another extension would result by going beyond
the~harmonic approximation and including all products of phonon operators. Method of
Bogoliubov\cite{bogol:1} could be applied to ``dangerous terms'' (divergent), omitted in 
Fr\"ohlich's~method. These questions will be dealt with in further investigations.
\appendix
\section{\label{a} The Fr\"ohlich's transformation}
Let us recall Fr\"ohlich's~method\cite{fr:1}, using in some details the~more elegant 
approach due to Davydov\cite{davydov}.

Consider the~electron-phonon Hamiltonian:
\begin{equation*}
	H_{\mathrm{e-ph}}=H_0+H_{\mathrm{int}}= 
	\sum_{\we{k}\sigma}\varepsilon_{\we{k}}a_{\we{k}\sigma}^\ast
	a_{\we{k}\sigma}+\sum_{\we{w}}\omega_{\we{w}}b_{\we{w}}^\ast
	b_{\we{w}}+i\sum_{\we{w}}D_{\we{w}}\left( 
	b_{\we{w}}\rho_{\we{w}}^\ast -b_{\we{w}}^\ast \rho_{\we{w}}\right),
\end{equation*}
where
\begin{equation*}
     	\rho_{\we{w}}=\sum_{\we{k}\sigma}a_{\we{k}-\we{w}\sigma}^\ast 
	a_{\we{k}\sigma},
\end{equation*}
and~$a_{\we{k}\sigma}$ ($b_{\we{k}}$) are fermion (boson) operators. 
The~coupling~$D_{\we{w}}$ will be assumed small and~$\hslash\equiv 1$.

The~form of the~interaction term of~$H_{\mathrm{e-ph}}$ arises under a~number of 
assumptions: the~ions of the~lattice move collectively, the~coupling depends only 
on~$\we{w}$ and electrons interact only with longitudinal phonons for 
which~$\omega_{\we{w}}=ws$, $s$~denoting the~velocity of sound. Our interest is 
focused on the~behaviour of electrons, therefore variations of the~phonon spectrum will 
be accounted for only through~$s$.

Fr\"ohlich performed a~unitary transformation of~$H_{\mathrm{e-ph}}$ in order to 
eliminate (as far as possible) the~interaction term. The~transformed Hamiltonian is
\begin{equation}
	H=\mathrm{e}^{S^\ast}H_{\mathrm{e-ph}}\mathrm{e}^S=H_{\mathrm{e-ph}}-
	[S,H_{\mathrm{e-ph}}]+\frac{1}{2}[S,[S,H_{\mathrm{e-ph}}]]+\ldots\, ,
\label{app_h_trans}
\end{equation}
where
\begin{equation*}
	S=\sum_{\we{q}}S_{\we{q}}=\sum_{\we{q}}\left( \gamma_{\we{q}}^\ast 
	b_{\we{q}}^\ast -\gamma_{\we{q}} b_{\we{q}}\right)=-S^\ast, \qquad
	\gamma_{\we{q}}=\sum_{\we{k}}\phi(\we{k},\we{q})a_{\we{k}}^\ast 
	a_{\we{k}-\we{q}},
\end{equation*}
and the~unknown function~$\phi(\we{k},\we{q}): \mathbb{R}^3\times
\mathbb{R}^3\to\mathbb{C}^1$ is adjusted to achieve the~cancellation.

Collecting terms of the~same order in the~coupling~$D_{\we{w}}$, one obtains 
Fr\"ohlich's~expansion:
\begin{equation}
	H=H_0-\left( [S,H_0]-H_{\mathrm{int}}\right)+\left( \frac{1}{2}[S,[S,H_0]]-
	[S,H_{\mathrm{int}}]\right)+\ldots\,.
\label{trans_2nd_order}
\end{equation}

Subsequently, a~term which is a~combination of products, each with $f$~fermion operators 
and $b$~boson operators will be written as~$(f,b)$. Clearly, $f$~will always be even. 
For example, $H_0$~consists of terms~$(2,0)$ and~$(0,2)$.

The rhs of Eq.~\eqref{trans_2nd_order} expresses in terms of 
commutators~$[(f_1,b_1),(f_2,b_2)]$. One easily finds that
\begin{equation}
	[(f_1,b_1),(f_2,b_2)]=[f_1,f_2]b_1 b_2 + f_2 f_1 [b_1,b_2]=[f_1,f_2]b_2 b_1 + 
	f_1 f_2 [b_1,b_2].
\label{app_com}
\end{equation}
The~necessary commutators~$[f_1,f_2]$, $[b_1,b_2]$ are given in Appendix~\ref{b}.

According to Eq.~\eqref{app_h_trans}, the~transformation can be performed, given
commutators of the~form occurring in Eq.~\eqref{app_com} with the~first argument 
equal~$S$. The~latter is a~$(2,1)$~expression, hence
\begin{equation*}
	[S,(f,b)]=[(2,1),(f,b)]=(f,b+1)+(f+2,b-1),
\end{equation*}
by virtue of Eqs.~\eqref{app_com},~\eqref{com_4f}. Clearly~$(f,b-1)=0$ for~$b=0$.

One finds
\begin{equation*}
 	[S,H_0]=[(2,1),(2,0)]+[(2,1),(0,2)]=(2,1),
\end{equation*}
so the~term arising from~$H_0$ in first order, has the~same form as the one arising 
from~$H_{\mathrm{int}}$ in zeroth order. These terms are used to eliminate 
the~interaction in~$H_{\mathrm{e-ph}}$.

Similarly, for~$H_{\mathrm{int}}$
\begin{equation*}
	[S,H_{\mathrm{int}}]=[(2,1),(2,1)] = (2,2)+(4,0).
\end{equation*}
Fr\"ohlich additionally assumed that introduction of experimentally measured velocity 
of sound~$s$ allows to discard all terms containing~$b_{\we{w}}^2$, 
$(b_{\we{w}}^\ast)^2$, $b_{\we{w}}b_{\we{v}}$, $b_{\we{w}}^\ast b_{\we{v}}$  
if~$\we{w}\neq  \we{v}$ etc. In other words, all $(f,b)$~terms for~$b\ge 3$ and part 
of $(f,2)$~terms are neglected. As a~consequence, the~expansion~\eqref{app_h_trans} up 
to 2nd~order reduces to\cite{fr:1}:
\begin{equation}
	\left[S,H_0\right]=\sum_{\we{k}\we{q}}b_{\we{q}}\left(\varepsilon_{\we{k}}-
	\varepsilon_{\we{k}-\we{q}}-\omega_{\we{q}}\right)\phi(\we{k},\we{q}) 
	a_{\we{k}}^\ast a_{\we{k}-\we{q}}+c.c.,
\label{sh0}
\end{equation}
\begin{equation}
	[S,H_{\mathrm{int}}]=\sum_{\we{q}}iD_{\we{q}}\gamma_{\we{q}}\rho_{\we{q}}+ 
	\sum_{\we{k}\we{q}} iD_{\we{q}} b_{\we{q}}^\ast b_{\we{q}}
	\phi(\we{k},\we{q})\left( n_{\we{k}}-n_{\we{k}-\we{q}}\right) +
	c.c.,
\label{sh1}
\end{equation}
\begin{equation}
	[S,[S,H_0]]=\sum_{\we{k}\we{q}}\left(\varepsilon_{\we{k}-\we{q}}-
	\varepsilon_{\we{k}}+\omega_{\we{q}}\right) b_{\we{q}}^\ast 
	b_{\we{q}}|\phi(\we{k},\we{q})|^2 \left(n_{\we{k}}-n_{\we{k}-\we{q}}\right)
 	+\sum_{\we{k}\we{q}}\left(\varepsilon_{\we{k}-\we{q}}-\varepsilon_{\we{k}}+
	\omega_{\we{q}}\right) \phi(\we{k},\we{q}) a_{\we{k}}^\ast a_{\we{k}-\we{q}} 
	\gamma_{\we{q}}^\ast +c.c..
\label{ssh0}
\end{equation}
$\phi$ is now adjusted so as to minimize the~contribution of terms~$(2,1)$. This is
achieved by the~choice\cite{fr:1}
\begin{equation}
	\phi(\we{k},\we{q})=\frac{-iD_{\we{q}}}{\varepsilon_{\we{k}-\we{q}}-
	\varepsilon_{\we{k}}+\omega_{\we{q}}} \left( 1-\Delta(\we{k},\we{q})\right), 
\label{phi}
\end{equation}
where
\begin{equation}
	\Delta(\we{k},\we{q})=\begin{cases} 
		1 \,\text{,}& \, \text{if}\quad \left| \varepsilon_{\we{k}-\we{q}}-
		\varepsilon_{\we{k}}+\omega_{\we{q}}\right| < \Gamma_{\we{q}} \\
	        0 \, \text{,}& \, \text{if}\quad \left| \varepsilon_{\we{k}-\we{q}}-
		\varepsilon_{\we{k}} +\omega_{\we{q}}\right| \ge \Gamma_{\we{q}} \\
	        \end{cases}.
\label{Delta}
\end{equation}
$\Delta(\we{k},\we{q})$ is introduced to avoid divergence of
the~series~\eqref{app_h_trans} and $\Gamma_{\we{q}}$ is positive energy choosing for
convergence.

Eqs.~\eqref{sh0}--\eqref{phi} yield
\begin{equation}
\begin{split}
 	H&=\sum_{\we{k}}\varepsilon_{\we{k}}a_{\we{k}}^\ast a_{\we{k}}+
	\sum_{\we{w}}\omega_{\we{w}}b_{\we{w}}^\ast b_{\we{w}}
 	+\sum_{\we{w}\we{k}}b_{\we{w}}^\ast b_{\we{w}}\left( n_{\we{k}}-n_{\we{k}-
	\we{w}}\right)\left(\frac{1}{2} \left(\varepsilon_{\we{k-\we{w}}}-
	\varepsilon_{\we{k}}+\omega_{\we{w}}\right)|\phi(\we{k},\we{w})|^2-
 	iD_{\we{w}}\phi(\we{k},\we{w})+ c.c.\right)\\
 	&+i\sum_{\we{k}\we{w}}D_{\we{w}}\left( b_{\we{w}}a_{\we{k}}^\ast a_{\we{k}-
	\we{w}}-b_{\we{w}}^\ast a_{\we{k}-\we{w}}^\ast a_{\we{k}}\right)\Delta(\we{k},
	\we{w})
 	-\frac{1}{2} \sum_{\we{k}\we{q}\we{w}}\frac{D_{\we{w}}^2\left( 1+\Delta(\we{k},
	\we{w})\right)\left( 1-\Delta(\we{q},\we{w})\right)}{\varepsilon_{\we{q}-\we{w}}-
	\varepsilon_{\we{q}}+\omega_{\we{w}}} \left(a_{\we{k}}^\ast a_{\we{k}-
	\we{w}}a_{\we{q}-\we{w}}^\ast a_{\we{q}}+c.c.\right).\\
\end{split}
\label{preH_F}
\end{equation}

Discarding the~non-transformed part of the~initial interaction (4th~sum on the rhs) and 
taking the~average in phonon vacuum, one obtains the~Fr\"ohlich's~Hamiltonian:
\begin{equation}
	H_{\mathrm{F}}=\sum_{\we{k}}\varepsilon_{\we{k}}n_{\we{k}}-\frac{1}{2} 
	\sum_{\we{k}\we{q}\we{w}} \frac{D_{\we{w}}^2 \left( 1+\Delta(\we{k},\we{w}) 
	\right)\left( 1-\Delta(\we{q},\we{w}) \right)}{\varepsilon_{\we{q}-\we{w}}-
	\varepsilon_{\we{q}}+\omega_{\we{w}}}\left( a_{\we{k}}^\ast 
	a_{\we{k}-\we{w}}a_{\we{q}-\we{w}}^\ast a_{\we{q}} +c.c.\right).
	\label{HF}
\end{equation}
The~second term represents an~effective interaction between electrons dressed in 
the~phonon field. 
If~$\varepsilon_{\we{q}-\we{w}}-\varepsilon_{\we{q}}+\omega_{\we{w}}>0$, this interaction 
is attractive.
\section{\label{c}Calculation of $g_{\we{k}\we{k}}$} 
Our objective is to find the~explicit form of the~following expression:
\begin{equation}
    	g_{\we{k}\we{k}}=-16\omega_{\we{k}}|\Lambda_{\we{k}\we{k}}|^2 
    	+ 4 \Re \Theta_{\we{k}\we{k}}.
\label{gkk_app}
\end{equation}

Taking into account Eq.~\eqref{Theta}, we have
\begin{equation}
\begin{split}
	\Theta_{\we{k}\we{k}}&=\omega_{\we{k}}\bigl( \chi^\ast(\we{k},\we{k},2\we{k},
	2\we{k})+\varphi^\ast(-\we{k},-\we{k},0,2\we{k})\bigr)\\
  	& \times \Bigl( \chi(3\we{k},\we{k},2\we{k},2\we{k})+\chi(\we{k},\we{k},0,2\we{k})
	+\varphi(\we{k},-\we{k},0,2\we{k})+\varphi(-\we{k},-\we{k},-2\we{k},
  	2\we{k})\Bigr)\\
	&=\omega_{\we{k}}\chi^\ast(\we{k},\we{k},2\we{k},2\we{k})\bigl( \chi(3\we{k},
  	\we{k},2\we{k},2\we{k})+\varphi(-\we{k},-\we{k},-2\we{k},2\we{k})\bigr),\\
\end{split}
\label{Thetakk}
\end{equation}
because
$\varphi^\ast(-\we{k},-\we{k},0,2\we{k})=\chi(\we{k},\we{k},0,2\we{k})=\varphi(\we{k}, 
-\we{k},0,2\we{k})=0$, as can be seen from the~explicit form e.g., of the~first of
these functions (see Eq.~\eqref{varphi}):
\begin{equation*}
\begin{split}
	\varphi^\ast(-\we{k},-\we{k},0,2\we{k})&=\frac{-iD_0^2
	D_{2\we{k}}}{6\omega_{\we{k}}^2 \omega_0} \left( 1 -
	\hat{\Delta}(-\we{k},-\we{k},0,2\we{k})\right)\left( 1-\Delta(\we{k},2\we{k})
	\right)\\
	&\times \Bigl\{ \left( 1-
	\Delta(-\we{k},0)\right) \left( -4 + \Delta(\we{k},0)\right)+ \left( 1- 
	\Delta(\we{k},0)\right) \left( 4 - \Delta(-\we{k},0)\right)\Bigr\} = 0,\\
\end{split}
\end{equation*}
because $\Delta(-\we{k},0)=\Delta(\we{k},0)=1$, for all $\we{k}$ 
(see Eq.~\eqref{Delta}). Similarly for~$\chi(\we{k},\we{k},0,2\we{k})$ and $\varphi(\we{k},-\we{k},0,2\we{k})$.

Similarly, from Eq.~\eqref{Lambda}, we get
\begin{equation}
	\Lambda_{\we{k}\we{k}}=\chi^\ast(3\we{k},\we{k},2\we{k},4\we{k})+\varphi^\ast(-
	\we{k},-3\we{k},-2\we{k},4\we{k}).
\label{Lambdakk}
\end{equation}

On the grounds of Eq.~\eqref{chi}, we obtain
\begin{equation}
\begin{split}
	&\chi^\ast(\we{k},\we{k},2\we{k},2\we{k})=\frac{-
	i\sqrt{2}D_{\we{k}}^3}{12\omega_{k}} 
	\left( 1-\tilde{\Delta}(\we{k},\we{k},2\we{k},2\we{k})\right)
	\left\{ \frac{3\left(1-
	\Delta(\we{k},2\we{k})\right)}{\omega_{\we{k}}^2}\right.\\
	& - \left.\frac{\left(1- \Delta(\we{k},2\we{k})\right) \left( 1- \Delta(-\we{k},
	2\we{k})\right)}{2\omega_{\we{k}}\left(	4\varepsilon_{\we{k}}+
	\omega_{\we{k}}\right)}
	\left( 2+\Delta(-\we{k},2\we{k})\right)-\frac{\left(1-\Delta(-\we{k},
	2\we{k})\right)}{2\left(4\varepsilon_{\we{k}}+\omega_{\we{k}}\right)^2}\left(4-
	\Delta(\we{k},2\we{k})\right) \right\}.\\
\end{split}
\label{chi1}
\end{equation}
The~functions $\tilde{\Delta}$ and $\Delta$ are equal $0$ or $1$, depending on their
argument (see Eqs.~\eqref{Delta_tilde}, \eqref{Delta}), so we obtain several conditions. 
The~most important one is
\begin{equation}
	\tilde{\Delta}(\we{k},\we{k},2\we{k},2\we{k})=0 \iff \omega_{\we{k}} >
	\tilde{\Gamma}_{\we{k}} \iff b > c,
\label{war1_1}
\end{equation}
and~$\chi^\ast(\we{k},\we{k},2\we{k},2\we{k})=0$ otherwise. Other
conditions are not so strong, because they do~not destroy all of parts
of Eq.~\eqref{chi1}. We have
\begin{equation}
	\Delta(\we{k},2\we{k})=0 \iff \omega_{\we{k}} > \Gamma_{\we{k}} \iff b > d,
\label{war1_2}
\end{equation}
\begin{equation}
	\Delta(-\we{k},2\we{k})=0 \iff |4\varepsilon_{\we{k}}+\omega_{\we{k}}| >
	\Gamma_{\we{k}} \iff k >\frac{d-b}{4a}.
\label{war1_3} 
\end{equation}
Taking into account Eqs.~\eqref{war1_1}, \eqref{war1_2} and \eqref{war1_3} we get
\begin{itemize}
	\item If $b \le c$, then $\chi^\ast(\we{k},\we{k},2\we{k},2\we{k})=0$.
        \item If $b > c$ and $b > d$, then 
		\begin{equation*}
		\chi^\ast(\we{k},\we{k},2\we{k},2\we{k})=\frac{-i\sqrt{2}D_{\we{k}}^3
		\varepsilon_{\we{k}}\left( 5\omega_{\we{k}}+12\varepsilon_{\we{k}} 
		\right)}{3\omega_{\we{k}}^3 \left( \omega_{\we{k}}+4\varepsilon_{\we{k}} 
		\right)^2}.
		\end{equation*}
	\item If $b > c$, $b \le d$ and $k \in \left( 0 , \frac{d-b}{4a} \right] $, then
	$\chi^\ast(\we{k},\we{k},2\we{k},2\we{k})=0$.
	\item If $b > c$, $b \le d$ and $k \in \left( \frac{d-b}{4a} , \infty \right)$, 
	then 
		\begin{equation*}
		\chi^\ast(\we{k},\we{k},2\we{k},2 \we{k})= 
		\frac{i\sqrt{2}D_{\we{k}}^3}{8 \omega_{\we{k}} \left( 
		\omega_{\we{k}}+4\varepsilon_{\we{k}}\right)^2}.
		\end{equation*}
\end{itemize}

Proceeding similarly, we get the~form of other functions occurring in
Eqs.~\eqref{Thetakk} and~\eqref{Lambdakk}. For~$\chi(3\we{k},\we{k},2\we{k},2\we{k})$,  
we have:
\begin{itemize}
	\item If $b \le c$, then $\chi(3\we{k},\we{k},2\we{k},2\we{k})=0$.
	\item If $b > c$,  $b > d$ and $k\in \left( 0 , \frac{b-d}{4a} \right) \cup
		\left( \frac{b+d}{4a} , \infty \right)$, then
                \begin{equation*}
		\chi(3\we{k},\we{k},2\we{k},2\we{k})=\frac{i\sqrt{2}D_{\we{k}}^3
                \varepsilon_{\we{k}}\left( 5\omega_{\we{k}}-12\varepsilon_{\we{k}} 
                \right)}{3\omega_{\we{k}}^3 \left( \omega_{\we{k}}-4\varepsilon_{\we{k}} 
                \right)^2}.
	        \end{equation*}
	\item If $b > c$, $b > d$ and $k\in \left[ \frac{b-d}{4a} , \frac{b+d}{4a}
		\right]$, then 
		\begin{equation*}
		\chi(3\we{k},\we{k},2\we{k},2\we{k})=-\frac{i\sqrt{2}D_{\we{k}}^3}{4 
		\omega_{\we{k}}^3}.
		\end{equation*}
        \item If $b > c$, $b \le d$ and $k \in \left[ 0 , \frac{b+d}{4a} 
		\right]$, then $\chi(3\we{k},\we{k},2\we{k},2\we{k})=0$.
	\item If $b > c$, $b \le d $ and $k\in \left( \frac{b+d}{4a} , \infty 
		\right)$, then
		\begin{equation*}
                \chi(3\we{k},\we{k},2\we{k},2\we{k})=\frac{i\sqrt{2}D_{\we{k}}^3}{8 
		\omega_{\we{k}} \left( \omega_{\we{k}}-4\varepsilon_{\we{k}} 
                \right)^2}.
                \end{equation*}
\end{itemize}

For~$\varphi(-\we{k},-\we{k},-2\we{k},2\we{k})$:
\begin{itemize}
	\item If $b \le e$, then $\varphi(-\we{k},-\we{k},-2\we{k},2\we{k})=0$.
	\item If $b > e$, $b > d$ and $k\in \left( 0 , \frac{b-d}{4a} \right)
		\cup \left( \frac{b+d}{4a} , \infty \right)$, then
                \begin{equation*}
		\varphi(-\we{k},-\we{k},-2\we{k},2\we{k})=\frac{i\sqrt{2}D_{\we{k}}^3
                \varepsilon_{\we{k}}\left( \omega_{\we{k}}+12\varepsilon_{\we{k}} 
                \right)}{3\omega_{\we{k}}^3 \left( \omega_{\we{k}}^2 - 16 
		\varepsilon_{\we{k}}^2 \right)}.
                \end{equation*}
	\item If $b > e$, $b > d$ and $k\in \left[ \frac{b-d}{4a} , \frac{b+d}{4a}
		\right]$, then
		\begin{equation*}
		\varphi(-\we{k},-\we{k},-2\we{k},2\we{k})=-\frac{i\sqrt{2} 
		D_{\we{k}}^3}{4\omega_{\we{k}}^3}.
		\end{equation*}
	\item If $b > e$, $b \le d$ and $k\in \left[ 0 , \frac{b+d}{4a}\right]$, then
		$\varphi(-\we{k},-\we{k},-2\we{k},2\we{k})=0$.
	\item If $b > e$, $b \le d$ and $k\in \left[ \frac{b+d}{4a} , \infty \right]$, 
		then
		\begin{equation*}
                \varphi(-\we{k},-\we{k},-2\we{k},2\we{k})=\frac{i\sqrt{2}
		D_{\we{k}}^3}{8\omega_{\we{k}}\left( \omega_{\we{k}}^2 - 16
		\varepsilon_{\we{k}}^2 \right)}.
		\end{equation*}
\end{itemize}

For~$\chi^\ast(3\we{k},\we{k},2\we{k},4\we{k})$:
\begin{itemize}
	\item If $b \le c$, then $\chi^\ast(3\we{k},\we{k},2\we{k},4\we{k})=0$.
	\item If $b > c$, $b < d / 3 $ and $ k \in \left( 0 , \frac{d-b}{2a}
		\right]$, then $\chi^\ast(3\we{k},\we{k},2\we{k},4\we{k})=0$.
	\item If $b > c$, $b < d / 3 $ and $ k \in \left( \frac{d-b}{2a} ,
		\frac{d+b}{2a} \right)$, then
                \begin{equation*}
		\chi^\ast(3\we{k},\we{k},2\we{k},4\we{k})=\frac{iD_{\we{k}}^3}{16 
		\omega_{\we{k}}\left( \omega_{\we{k}} + 2 \varepsilon_{\we{k}} \right) 
		\left( \omega_{\we{k}} +
		4\varepsilon_{\we{k}}\right)}:= \chi^{(1)}.
                \end{equation*}
	\item If $b > c$, $b < d / 3 $ and $ k \in \left( \frac{d+b}{2a} , \infty  
		\right)$, then
                \begin{equation*}
		\chi^\ast(3\we{k},\we{k},2\we{k},4\we{k})=\frac{-3iD_{\we{k}}^3
		\varepsilon_{\we{k}}}{4 \left( \omega_{\we{k}}^2 - 16
		\varepsilon_{\we{k}}^2 \right) \left( \omega_{\we{k}}^2 -
		4\varepsilon_{\we{k}}^2\right)}:= \chi^{(2)}.
		\end{equation*}
	\item If $b > c$, $ d / 3 \le b \le d $ and $ k \in \left( 0 , \frac{d-b}{2a}
                \right]$, then $\chi^\ast(3\we{k},\we{k},2\we{k},4\we{k})=0$.
	\item If $b > c$, $ d / 3 \le b \le d $ and $ k \in \left( \frac{d-b}{2a} ,
		\frac{d+b}{2a} \right]$, then
		$\chi^\ast(3\we{k},\we{k},2\we{k},4\we{k})=\chi^{(1)}$.
	\item If $b > c$, $ d / 3 \le b \le d $ and $ k \in \left( \frac{d+b}{2a} ,
		\infty \right)$, then
		$\chi^\ast(3\we{k},\we{k},2\we{k},4\we{k})=\chi^{(2)}$.
	\item If $b > c$, $ d < b \le 3d $ and $ k \in \left( 0 , \frac{b-d}{4a}
		\right) \cup \left( \frac{b+d}{2a}, \infty \right)$, then
		\begin{equation*}
                \chi^\ast(3\we{k},\we{k},2\we{k},4\we{k})=\frac{-iD_{\we{k}}^3
		\varepsilon_{\we{k}} \left( 7\omega_{\we{k}}^2 - 16
		\varepsilon_{\we{k}}^2 \right)}{6 \omega_{\we{k}}^2 \left( 
		\omega_{\we{k}}^2 - 16 \varepsilon_{\we{k}}^2 \right) \left(
		\omega_{\we{k}}^2 - 4\varepsilon_{\we{k}}^2\right)}:=
		\chi^{(3)}.
                \end{equation*}
	\item If $b > c$, $ d < b \le 3d $ and $ k \in \left[ \frac{b-d}{4a} ,
		\frac{b-d}{2a} \right)$, then
		\begin{equation*}
                \chi^\ast(3\we{k},\we{k},2\we{k},4\we{k})=\frac{-iD_{\we{k}}^3
                \left( -3\omega_{\we{k}}^2 + 40 \varepsilon_{\we{k}}^2 + 22
		\omega_{\we{k}} \varepsilon_{\we{k}} \right)}{48 \omega_{\we{k}}^2
		\left( \omega_{\we{k}}^2 - 4 \varepsilon_{\we{k}}^2 \right)
		\left( \omega_{\we{k}} + 4\varepsilon_{\we{k}}\right)}:=
                \chi^{(4)}.
                \end{equation*}
	\item If $b > c$, $ d < b \le 3d $ and $ k \in \left[ \frac{b-d}{2a} ,
		\frac{b+d}{2a}\right]$, then
		\begin{equation*}
                \chi^\ast(3\we{k},\we{k},2\we{k},4\we{k})=\frac{iD_{\we{k}}^3
                \left( 3\omega_{\we{k}} + 4 \varepsilon_{\we{k}}\right)}{24
		\omega_{\we{k}}^2 \left( \omega_{\we{k}} + 2 \varepsilon_{\we{k}}
		\right) \left( \omega_{\we{k}} + 4\varepsilon_{\we{k}}\right)}
		:=\chi^{(5)}.
                \end{equation*}
        \item If $b > c$, $  b > 3d $ and $ k \in \left( 0 ,
		\frac{b-d}{4a} \right) \cup \left( \frac{b+d}{4a},
		\frac{b-d}{2a} \right) \cup \left( \frac{b+d}{2a},
		\infty \right)$,
		then
		$\chi^\ast(3\we{k},\we{k},2\we{k},4\we{k})=\chi^{(3)}$.
        \item If $b > c$, $  b > 3d $ and $ k \in \left[ \frac{b-d}{4a} ,
		\frac{b+d}{4a} \right]$, then
                $\chi^\ast(3\we{k},\we{k},2\we{k},4\we{k})=\chi^{(4)}$.
        \item If $b > c$, $  b > 3d $ and $ k \in \left[ \frac{b-d}{2a},
                \frac{b+d}{2a} \right]$, then
                $\chi^\ast(3\we{k},\we{k},2\we{k},4\we{k})=\chi^{(5)}$.
\end{itemize}

For $\varphi^\ast(-\we{k},-3\we{k},-2\we{k},4\we{k})$:
\begin{itemize}
        \item If $b \le e$, then $\varphi^\ast(-\we{k},-3\we{k},-2\we{k},4\we{k})=0$.
	\item If $b > e$, $b < d / 3 $ and $ k \in \left( 0 , \frac{d-b}{2a}
		\right]$, then $\varphi^\ast(-\we{k},-3\we{k},-2\we{k},4\we{k})=0$.
        \item If $b > e$, $b < d / 3 $ and $ k \in \left( \frac{d-b}{2a} ,
                \frac{d+b}{2a} \right]$, then
                \begin{equation*}
		\varphi^\ast(-\we{k},-3\we{k},-2\we{k},4\we{k})=
		\frac{iD_{\we{k}}^3}{16\omega_{\we{k}}\left( \omega_{\we{k}} + 2
		\varepsilon_{\we{k}} \right)\left( \omega_{\we{k}} -
		4\varepsilon_{\we{k}}\right)}:= \varphi^{(1)}.
                \end{equation*}
        \item If $b > e$, $b < d / 3 $ and $ k \in \left( \frac{d+b}{2a} , \infty
                \right)$, then
                \begin{equation*}
                \varphi^\ast(-\we{k},-3\we{k},-2\we{k},4\we{k})=
		\frac{iD_{\we{k}}^3\varepsilon_{\we{k}}}{4 \left( \omega_{\we{k}}^2
		- 16\varepsilon_{\we{k}}^2 \right) \left( \omega_{\we{k}}^2
		-4\varepsilon_{\we{k}}^2\right)}:= \varphi^{(2)}.
                \end{equation*}
        \item If $b > e$, $ d / 3 \le b \le d $ and $ k \in \left( 0 , \frac{d+b}{4a}
                \right]$, then $\varphi^\ast(-\we{k},-3\we{k},-2\we{k},4\we{k})=0$.
        \item If $b > e$, $ d / 3 \le b \le d $ and $ k \in \left( \frac{d+b}{4a} ,
		\frac{d+b}{2a} \right]$, then
		$\varphi^\ast(-\we{k},-3\we{k},-2\we{k},4\we{k})=\varphi^{(1)}$.
        \item If $b > e$, $ d / 3 \le b \le d $ and $ k \in \left( \frac{d+b}{2a} ,
		\infty \right)$, then
                $\varphi^\ast(-\we{k},-3\we{k},-2\we{k},4\we{k})=\varphi^{(2)}$.
        \item If $b > e$, $ d < b \le 3d $ and $ k \in \left( 0 , \frac{b-d}{4a}
		\right) \cup \left[ \frac{b+d}{2a} , \infty \right)$, then 
                \begin{equation*}
                \varphi^\ast(-\we{k},-3\we{k},-2\we{k},4\we{k})=
		\frac{iD_{\we{k}}^3\varepsilon_{\we{k}}\left( \omega_{\we{k}}^2
		+ 16\varepsilon_{\we{k}}^2\right)}{6 \omega_{\we{k}}^2 \left( 
		\omega_{\we{k}}^2-4\varepsilon_{\we{k}}^2\right)\left(
		\omega_{\we{k}}^2 - 16\varepsilon_{\we{k}}^2 \right)}
		:= \varphi^{(3)}.
		\end{equation*}
        \item If $b > e$, $ d < b \le 3d $ and $ k \in \left[ \frac{b-d}{4a} ,
		\frac{b-d}{2a} \right)$, then
                \begin{equation*}
                \varphi^\ast(-\we{k},-3\we{k},-2\we{k},4\we{k})=
                \frac{-iD_{\we{k}}^3\left( 3 \omega_{\we{k}}^2 + 14
		\varepsilon_{\we{k}} \omega_{\we{k}} + 40 \varepsilon_{\we{k}}^2 
		\right)}{48 \omega_{\we{k}}^2 \left( \omega_{\we{k}}^2 - 4 
		\varepsilon_{\we{k}}^2 \right) \left( \omega_{\we{k}} +
		4\varepsilon_{\we{k}}\right)}:= \varphi^{(4)}.
                \end{equation*}
        \item If $b > e$, $ d < b \le 3d $ and $ k \in \left[ \frac{b-d}{2a} , 
                \frac{b+d}{4a} \right]$, then
                \begin{equation*}
                \varphi^\ast(-\we{k},-3\we{k},-2\we{k},4\we{k})=
		\frac{iD_{\we{k}}^3}{16 \omega_{\we{k}}^2 \left( \omega_{\we{k}} +
		2\varepsilon_{\we{k}}\right)}.
                \end{equation*}
        \item If $b > e$, $ d < b \le 3d $ and $ k \in \left( \frac{b+d}{4a} , 
                \frac{b+d}{2a} \right)$, then
                \begin{equation*}
                \varphi^\ast(-\we{k},-3\we{k},-2\we{k},4\we{k})=
		\frac{iD_{\we{k}}^3 \left( 3 \omega_{\we{k}} - 4 \varepsilon_{\we{k}}
		\right)}{24 \omega_{\we{k}}^2 \left( \omega_{\we{k}} + 2 
		\varepsilon_{\we{k}}\right) \left( \omega_{\we{k}} - 4
		\varepsilon_{\we{k}}\right)}:= \varphi^{(5)}.
                \end{equation*}
        \item If $b > e$, $ 3d < b $ and $ k \in \left( 0 , \frac{b-d}{4a} 
		\right) \cup \left( \frac{b+d}{4a} , \frac{b-d}{2a}
		\right) \cup \left( \frac{b+d}{2a} , \infty \right)$, then
                $\varphi^\ast(-\we{k},-3\we{k},-2\we{k},4\we{k})=\varphi^{(3)}$.
	\item If $b > e$, $ 3d < b $ and $ k \in \left[ \frac{b-d}{4a} ,
		\frac{b+d}{4a} \right]$, then
                $\varphi^\ast(-\we{k},-3\we{k},-2\we{k},4\we{k})=\varphi^{(4)}$.
        \item If $b > e$, $ 3d < b $ and $ k \in \left[ \frac{b-d}{2a} ,
                \frac{b+d}{2a} \right]$, then
                $\varphi^\ast(-\we{k},-3\we{k},-2\we{k},4\we{k})=\varphi^{(5)}$.
\end{itemize}

On the~basis of this results and Eq.~\eqref{gkk_app}, we get the~4-fermion 
interaction coupling~$g_{\we{k}\we{k}}$:
\begin{itemize}
	\item If $ b \le c$ and $ b \le e$, then $g_{\we{k}\we{k}}=0$.
	\item If $ b \le c$, $ b > e$, $ b < d/3$ and $k \in \left( 0 ,
		\frac{d-b}{2a} \right] $, then $g_{\we{k}\we{k}}=0$.
        \item If $ b \le c$, $ b > e$, $ b < d/3$ and $k \in \left(
		\frac{d-b}{2a} , \frac{d+b}{2a} \right]$, then 
		\begin{equation*}
		g_{\we{k}\we{k}}=
		\frac{-D_{\we{k}}^6}{16 \omega_{\we{k}} \left( \omega_{\we{k}} + 2
		\varepsilon_{\we{k}} \right)^2 \left( \omega_{\we{k}} - 4
		\varepsilon_{\we{k}} \right)^2} := g_{\we{k}\we{k}}^{(1)} < 0.
		\end{equation*}
	\item If $ b \le c$, $ b > e$, $ b < d/3$ and $k \in \left( \frac{d+b}{2a} ,
		\infty \right)$, then
	        \begin{equation*}
                g_{\we{k}\we{k}}=
		\frac{-D_{\we{k}}^6 \omega_{\we{k}} \varepsilon_{\we{k}}^2}{\left( 
		\omega_{\we{k}}^2 - 4 \varepsilon_{\we{k}}^2 \right)^2 \left( 
		\omega_{\we{k}}^2 - 16 \varepsilon_{\we{k}}^2 \right)^2} :=
		g_{\we{k}\we{k}}^{(2)}< 0.
        	\end{equation*}
        \item If $ b \le c$, $ b > e$, $ d/3 \le b \le d$ and $k \in \left( 0 , 
		\frac{d+b}{4a} \right]$, then $g_{\we{k}\we{k}}=0$.
        \item If $ b \le c$, $ b > e$, $ d/3 \le b \le d$ and $k \in \left(
		\frac{d+b}{4a} , \frac{d+b}{2a} \right]$, then
		$g_{\we{k}\we{k}}=g_{\we{k}\we{k}}^{(1)}$.
        \item If $ b \le c$, $ b > e$, $ d/3 \le b \le d$ and $k \in \left(
		\frac{d+b}{2a} , \infty \right)$, then
		$g_{\we{k}\we{k}}=g_{\we{k}\we{k}}^{(2)}$.
        \item If $ b \le c$, $ b > e$, $ d <  b \le 3d$ and $k \in \left( 0 ,
		\frac{b-d}{4a} \right) \cup \left[ \frac{b+d}{2a} , \infty \right)$, then 
	        \begin{equation*}
                g_{\we{k}\we{k}}=
		-\frac{4 D_{\we{k}}^6 \varepsilon_{\we{k}}^2 \left( \omega_{\we{k}}^2
		+ 16 \varepsilon_{\we{k}}^2 \right)^2}{9 \omega_{\we{k}}^3 \left(
		\omega_{\we{k}}^2 - 4 \varepsilon_{\we{k}}^2 \right)^2 \left(
		\omega_{\we{k}}^2 - 16 \varepsilon_{\we{k}}^2 \right)^2}:=
		g_{\we{k}\we{k}}^{(3)} < 0.
	        \end{equation*}
	\item If $ b \le c$, $ b > e$, $ d <  b \le 3d$ and $k \in \left[
		\frac{b-d}{4a} ,  \frac{b-d}{2a} \right)$, then
        	\begin{equation*}
                g_{\we{k}\we{k}}=
                \frac{-D_{\we{k}}^6 \left( 3 \omega_{\we{k}}^2 + 14
		\varepsilon_{\we{k}} \omega_{\we{k}} + 40 \varepsilon_{\we{k}}^2
		\right)^2}{144 \omega_{\we{k}}^3 \left( \omega_{\we{k}}^2 - 4
		\varepsilon_{\we{k}}^2 \right)^2 \left( \omega_{\we{k}} + 4
		\varepsilon_{\we{k}} \right)^2}:= g_{\we{k}\we{k}}^{(4)} < 0.
	\end{equation*}
	\item If $ b \le c$, $ b > e$, $ d <  b \le 3d$ and $k \in \left[
		\frac{b-d}{2a} , \frac{b+d}{4a} \right]$, then
	        \begin{equation*}
                g_{\we{k}\we{k}}=
		-\frac{D_{\we{k}}^6}{16 \omega_{\we{k}}^3 \left( \omega_{\we{k}} + 2   
		\varepsilon_{\we{k}} \right)^2} < 0.
	        \end{equation*}
        \item If $ b \le c$, $ b > e$, $ d <  b \le 3d$ and $k \in \left(
		\frac{b+d}{4a} , \frac{b+d}{2a} \right)$, then
	        \begin{equation*}
                g_{\we{k}\we{k}}=
		\frac{-D_{\we{k}}^6 \left( 3 \omega_{\we{k}} - 4
		\varepsilon_{\we{k}} \right)^2}{36 \omega_{\we{k}}^3 \left( 
		\omega_{\we{k}} + 2 \varepsilon_{\we{k}} \right)^2 \left( 
		\omega_{\we{k}} - 4 \varepsilon_{\we{k}} \right)^2}:=
		g_{\we{k}\we{k}}^{(5)} < 0.
        \end{equation*}
	\item If $ b \le c$, $ b > e$, $ 3d <  b $ and $k \in \left( 0 ,
		\frac{b-d}{4a} \right) \cup \left( \frac{b+d}{4a} , \frac{b-d}{2a} 
		\right) \cup \left( \frac{b+d}{2a} , \infty \right)$, then
		$g_{\we{k}\we{k}}=g_{\we{k}\we{k}}^{(3)}$.
	\item If $ b \le c$, $ b > e$, $ 3d <  b $ and $k \in \left[  \frac{b-d}{4a} ,
		\frac{b+d}{4a} \right]$, then
		$g_{\we{k}\we{k}}=g_{\we{k}\we{k}}^{(4)}$.
        \item If $ b \le c$, $ b > e$, $ 3d <  b $ and $k \in \left[ \frac{b-d}{2a} ,
		\frac{b+d}{2a} \right]$, then
		$g_{\we{k}\we{k}}=g_{\we{k}\we{k}}^{(5)}$.
        \item If $ b > c$, $ b \le e$, $ b < d/3$ and $k \in \left( 0 , \frac{b+d}{4a} 
		\right]$, then $g_{\we{k}\we{k}}=0$.
        \item If $ b > c$, $ b \le e$, $ b < d/3$ and $k \in \left( \frac{b+d}{4a} ,
		\frac{d-b}{2a} \right]$, then
        	\begin{equation*}
                g_{\we{k}\we{k}}=
		\frac{-D_{\we{k}}^6}{8 \omega_{\we{k}} \left( \omega_{\we{k}}^2
		-16 \varepsilon_{\we{k}}^2 \right)^2} < 0.
	        \end{equation*}
        \item If $ b > c$, $ b \le e$, $ b < d/3$ and $k \in \left( \frac{d-b}{2a} ,
		\frac{b+d}{2a} \right]$, then
        	\begin{equation*}
                g_{\we{k}\we{k}}=
		\frac{-3 D_{\we{k}}^6 \left( \omega_{\we{k}}^2 + 8
		\varepsilon_{\we{k}}^2 \right)}{16 \omega_{\we{k}} \left(
		\omega_{\we{k}} + 2 \varepsilon_{\we{k}} \right)^2 \left( 
		\omega_{\we{k}}^2 - 16 \varepsilon_{\we{k}}^2 \right)^2}:=
		g_{\we{k}\we{k}}^{(6)} < 0.
	        \end{equation*}
        \item If $ b > c$, $ b \le e$, $ b < d/3$ and $k \in \left( \frac{d+b}{2a} ,
		\infty \right)$, then 
        	\begin{equation*}
                g_{\we{k}\we{k}}=
                \frac{-D_{\we{k}}^6 \left( \omega_{\we{k}}^4 + 64 \omega_{\we{k}}^2 
		\varepsilon_{\we{k}}^2 + 16 \varepsilon_{\we{k}}^4 \right)}{8
		\omega_{\we{k}} \left( \omega_{\we{k}}^2 - 16 \varepsilon_{\we{k}}^2
		\right)^2 \left( \omega_{\we{k}}^2 - 4 \varepsilon_{\we{k}}^2 
		\right)^2}:= g_{\we{k}\we{k}}^{(7)} < 0.
	        \end{equation*}
        \item If $ b > c$, $ b \le e$, $ d/3 \le b \le d$ and $k \in \left( 0 ,
		\frac{d-b}{2a} \right]$, then $g_{\we{k}\we{k}}=0$.
        \item If $ b > c$, $ b \le e$, $ d/3 \le b \le d$ and $k \in \left(
		\frac{d-b}{2a} , \frac{d+b}{4a}\right]$, then
        	\begin{equation*}
                g_{\we{k}\we{k}}=
		\frac{-D_{\we{k}}^6}{16 \omega_{\we{k}} \left( \omega_{\we{k}} + 2
		\varepsilon_{\we{k}} \right)^2 \left( \omega_{\we{k}} + 4
		\varepsilon_{\we{k}} \right)^2}:= g_{\we{k}\we{k}}^{(8)} < 0.
	        \end{equation*}
        \item If $ b > c$, $ b \le e$, $ d/3 \le b \le d$ and $k \in \left(
		\frac{d+b}{4a} , \frac{d+b}{2a} \right]$, then 
		$ g_{\we{k}\we{k}} = g_{\we{k}\we{k}}^{(6)}$.
        \item If $ b > c$, $ b \le e$, $ d/3 \le b \le d$ and $k \in \left(
		\frac{d+b}{2a} , \infty \right)$, then
		$ g_{\we{k}\we{k}} = g_{\we{k}\we{k}}^{(7)}$.
        \item If $ b > c$, $ b \le e$, $ d < b \le 3d$ and $k \in \left( 0 ,
		\frac{b-d}{4a} \right) \cup \left( \frac{b+d}{2a} , \infty \right)$, then
        	\begin{equation*}
                g_{\we{k}\we{k}}=
		-\frac{4 D_{\we{k}}^6 \varepsilon_{\we{k}}^2 \left( -
		\omega_{\we{k}}^6 +
		464 \omega_{\we{k}}^4 \varepsilon_{\we{k}}^2 - 2848 \omega_{\we{k}}^2
		\varepsilon_{\we{k}}^4 + 4608 \varepsilon_{\we{k}}^6 \right)}{9
		\omega_{\we{k}}^5 \left( \omega_{\we{k}}^2 - 4 \varepsilon_{\we{k}}^2
		\right)^2 \left( \omega_{\we{k}}^2 - 16 \varepsilon_{\we{k}}^2
		\right)^2} := g_{\we{k}\we{k}}^{(9)}.
        	\end{equation*}
	We see that~$g_{\we{k}\we{k}}^{(9)} < 0$ iff
	\begin{equation*}
	\omega_{\we{k}}^6 + 464 \omega_{\we{k}}^4 \varepsilon_{\we{k}}^2 -
	2848 \omega_{\we{k}}^2 \varepsilon_{\we{k}}^4 + 4608 
	\varepsilon_{\we{k}}^6 > 0.
	\end{equation*}
	Taking into account~$\omega_{\we{k}}=bk$ and~$\varepsilon_{\we{k}}=ak^2$, we
	obtain a~3rd order algebraic inequality 
	\begin{equation*}
	4608 l^3 - 2848 x l^2 + 464 x^2 l - x^3 > 0,
	\end{equation*}
	for~$l=a^2 k^2$, $x=b^2$. This can be easily solved\footnote{this and following 
	inequalities are solved by \emph{Mathematica} package}---there exist only one 
	real root~$l_0^{(9)} \approx 0, 002 x$, so 
	\begin{equation*}
		g_{\we{k}\we{k}}^{(9)} < 0 \quad \text{iff} \quad
		k > k_0^{(9)} \approx 0.047 b /a < b/4a.
	\end{equation*}. 
	\item If $ b > c$, $ b \le e$, $ d < b \le 3d$ and $k \in \left[
		\frac{b-d}{4a} , \frac{b-d}{2a} \right)$, then
        \begin{equation*}
		\begin{split}
		g_{\we{k}\we{k}} & =
		-\frac{D_{\we{k}}^6}{144 \omega_{\we{k}}^5 \left(
                \omega_{\we{k}}^2 - 4 \varepsilon_{\we{k}}^2 \right)^2 \left(
                \omega_{\we{k}} + 4 \varepsilon_{\we{k}} \right)^2} \\
		& \times \left( 9 \omega_{\we{k}}^6 + 348
		\omega_{\we{k}}^5 \varepsilon_{\we{k}} + 1396 \omega_{\we{k}}^4
		\varepsilon_{\we{k}}^2 - 2080 \omega_{\we{k}}^3
		\varepsilon_{\we{k}}^3 - 7616 \omega_{\we{k}}^2 
		\varepsilon_{\we{k}}^4 + 7680 \omega_{\we{k}} \varepsilon_{\we{k}}^5
		+ 18432 \varepsilon_{\we{k}}^6 \right):= g_{\we{k}\we{k}}^{(10)}.\\
	\end{split}
	\end{equation*}
	It~turns out that~$g_{\we{k}\we{k}}^{(10)} < 0$ for all~$k>0$, because ($l=a k$) 
	the~equation 
	\begin{equation*}
		18432 l^6 + 7680 b l^5 - 7616 b^2 l^4 - 2080 b^3 l^3 +
		1396 b^4 l^2 + 348 b^5 l + 9 b^6 = 0 
	\end{equation*}
	has only two real roots and both are negative: $l_1^{(10)} \approx -0.208 b$ 
	and~$l_1^{(10)}=-0.029 b$.
	\item If $ b > c$, $ b \le e$, $ d < b \le 3d$ and $k \in \left[
                \frac{b-d}{2a} , \frac{b+d}{4a} \right]$, then
        \begin{equation*}
                g_{\we{k}\we{k}}=
		-\frac{D_{\we{k}}^6 \left( 9 \omega_{\we{k}}^4 + 144
		\omega_{\we{k}}^3 \varepsilon_{\we{k}} + 784 \omega_{\we{k}}^2 
		\varepsilon_{\we{k}}^2 + 1632 \omega_{\we{k}} \varepsilon_{\we{k}}^3
		+ 1152 \varepsilon_{\we{k}}^4 \right)}{36 \omega_{\we{k}}^5
		\left( \omega_{\we{k}} + 2 \varepsilon_{\we{k}} \right)^2
		\left( \omega_{\we{k}} + 4 \varepsilon_{\we{k}} \right)^2} < 0.
        \end{equation*}
        \item If $ b > c$, $ b \le e$, $ d < b \le 3d$ and $k \in \left(
                \frac{b+d}{4a} , \frac{b+d}{2a} \right]$, then
        \begin{equation*}
		\begin{split}
                g_{\we{k}\we{k}} & =
		-\frac{D_{\we{k}}^6}{36 \omega_{\we{k}}^5 \left( \omega_{\we{k}} + 2
		\varepsilon_{\we{k}} \right)^2 \left( \omega_{\we{k}}^2 - 16
		\varepsilon_{\we{k}}^2 \right)^2} \\
		& \times \left( 9 \omega_{\we{k}}^6 - 48 \omega_{\we{k}}^5
		\varepsilon_{\we{k}} -832 \omega_{\we{k}}^4 \varepsilon_{\we{k}}^2 -
		2944 \omega_{\we{k}}^3 \varepsilon_{\we{k}}^3 
		+ 1664 \omega_{\we{k}}^2 \varepsilon_{\we{k}}^4 + 18432 
		\omega_{\we{k}} \varepsilon_{\we{k}}^5 +18432 \varepsilon_{\we{k}}^6 
		\right) := g_{\we{k}\we{k}}^{(11)}.\\
	\end{split}
        \end{equation*}
	$g_{\we{k}\we{k}}^{(11)}< 0$ iff $k \in \left( 0 , 0.073 b/a
	\right) \cup \left( 0.412 b/a , \infty \right)$.
	\item If $ b > c$, $ b \le e$, $  b > 3d$ and $k \in \left( 0 , \frac{b-d}{4a}
		\right) \cup \left( \frac{b+d}{4a} , \frac{b-d}{2a} \right) \cup \left(
		\frac{b+d}{2a} , \infty \right)$, then
		$g_{\we{k}\we{k}}=g_{\we{k}\we{k}}^{(9)}$.
	\item If $ b > c$, $ b \le e$, $  b > 3d$ and $k \in \left[ \frac{b-d}{4a} ,
		\frac{b+d}{4a} \right]$, then 
        	$g_{\we{k}\we{k}}=g_{\we{k}\we{k}}^{(10)}$.
	\item If $ b > c$, $ b \le e$, $  b > 3d$ and $k \in \left[ \frac{b-d}{2a} ,
		\frac{b+d}{2a} \right]$, then
                $g_{\we{k}\we{k}}=g_{\we{k}\we{k}}^{(11)}$.
        \item If $ b > c$, $ b > e$, $  b < d/3$ and $k \in \left( 0 , \frac{b+d}{4a}
		\right]$, then $g_{\we{k}\we{k}}=0$.
        \item If $ b > c$, $ b > e$, $  b < d/3$ and $k \in \left( \frac{b+d}{4a} ,
		\frac{d-b}{2a} \right)$, then
        \begin{equation*}
                g_{\we{k}\we{k}}=
		-\frac{D_{\we{k}}^6}{4 \left( \omega_{\we{k}} + 4
		\varepsilon_{\we{k}} \right)^3 \left( \omega_{\we{k}} - 4
		\varepsilon_{\we{k}} \right)^2} < 0.
        \end{equation*}
        \item If $ b > c$, $ b > e$, $  b < d/3$ and $k \in \left[ \frac{d-b}{2a} ,
		\frac{b+d}{2a} \right]$, then
        \begin{equation*}
                g_{\we{k}\we{k}}=
		-\frac{D_{\we{k}}^6 \left( \omega_{\we{k}}^2 + 4 \omega_{\we{k}}
		\varepsilon_{\we{k}} + 2 \varepsilon_{\we{k}}^2 \right)}{2 \left( 
		\omega_{\we{k}}^2 - 16 \varepsilon_{\we{k}}^2 \right)^2 \left(
		\omega_{\we{k}} + 2 \varepsilon_{\we{k}} \right)^2 \left(
		\omega_{\we{k}} + 4 \varepsilon_{\we{k}} \right)}:=
		g_{\we{k}\we{k}}^{(12)} < 0.
        \end{equation*}
        \item If $ b > c$, $ b > e$, $  b < d/3$ and $k \in \left( \frac{d+b}{2a} ,
		\infty \right)$, then
        \begin{equation*}
                g_{\we{k}\we{k}}=
                -\frac{D_{\we{k}}^6 \left( \omega_{\we{k}}^4 - 4 \omega_{\we{k}}^2
		\varepsilon_{\we{k}}^2 + 16 \omega_{\we{k}} \varepsilon_{\we{k}}^3 +
		16 \varepsilon_{\we{k}}^4 \right)}{4 \left( \omega_{\we{k}}^2 - 16
		\varepsilon_{\we{k}}^2 \right)^2 \left( \omega_{\we{k}}^2 - 4
		\varepsilon_{\we{k}}^2 \right)^2 \left( \omega_{\we{k}} + 4
		\varepsilon_{\we{k}} \right)}:= g_{\we{k}\we{k}}^{(13)}.
        \end{equation*}
	$g_{\we{k}\we{k}}^{(13)} < 0$ for all $k >0$.
        \item If $ b > c$, $ b > e$, $ d/3 \le b \le d$ and $k \in \left( 0 ,
		\frac{d-b}{2a} \right]$, then $g_{\we{k}\we{k}}=0$.
        \item If $ b > c$, $ b > e$, $ d/3 \le b \le d$ and $k \in \left(
		\frac{d-b}{2a} , \frac{d+b}{4a} \right]$, then
		$g_{\we{k}\we{k}}=g_{\we{k}\we{k}}^{(8)}$.
	\item If $ b > c$, $ b > e$, $ d/3 \le b \le d$ and $k \in \left(
		\frac{d+b}{4a} , \frac{d+b}{2a} \right]$, then
		$g_{\we{k}\we{k}}=g_{\we{k}\we{k}}^{(12)}$.
        \item If $ b > c$, $ b > e$, $ d/3 \le b \le d$ and $k \in \left(
		\frac{d+b}{2a}, \infty \right)$, then
		$g_{\we{k}\we{k}}=g_{\we{k}\we{k}}^{(13)}$.
	\item If $ b > c$, $ b > e$, $ d < b \le 3d$ and $k \in \left( 0 ,
		\frac{b-d}{4a} \right)$, then 
        \begin{equation*}
		\begin{split}
                g_{\we{k}\we{k}} & =
		-\frac{32 D_{\we{k}}^6 \varepsilon_{\we{k}}^2}{9 \omega_{\we{k}}^5
		\left( \omega_{\we{k}}^2 - 16 \varepsilon_{\we{k}}^2 \right)^2 \left(
		\omega_{\we{k}}^2 - 4 \varepsilon_{\we{k}}^2 \right)^2 \left(
		\omega_{\we{k}} + 4 \varepsilon_{\we{k}} \right)}\\		
		& \times \left( -3 \omega_{\we{k}}^7 - 20 \omega_{\we{k}}^6 
		\varepsilon_{\we{k}} + 84 \omega_{\we{k}}^5 \varepsilon_{\we{k}}^2 + 
		400 \omega_{\we{k}}^4 \varepsilon_{\we{k}}^3 
		- 568 \omega_{\we{k}}^3 
		\varepsilon_{\we{k}}^4 - 2400 \omega_{\we{k}}^2 
		\varepsilon_{\we{k}}^5 + 1152 \omega_{\we{k}} 
		\varepsilon_{\we{k}}^6 + 4608 \varepsilon_{\we{k}}^7
		\right) := g_{\we{k}\we{k}}^{(14)}.\\
	\end{split}
        \end{equation*}
	$g_{\we{k}\we{k}}^{(14)} < 0 $ iff $k \in \left( 0.268 b/a ,
	\infty \right)$.
	\item If $ b > c$, $ b > e$, $ d < b \le 3d$ and $k \in \left[ \frac{b-d}{4a}
		, \frac{b-d}{2a} \right)$, then
        \begin{equation*}
                g_{\we{k}\we{k}}=
		-\frac{16 D_{\we{k}}^6 \varepsilon_{\we{k}} \left( 60
		\omega_{\we{k}}^5 + 225 \omega_{\we{k}}^4 \varepsilon_{\we{k}} - 120
		\omega_{\we{k}}^3 \varepsilon_{\we{k}}^2 -752 \omega_{\we{k}}^2
		\varepsilon_{\we{k}}^3 + 960 \omega_{\we{k}} \varepsilon_{\we{k}}^4 +
		2304 \varepsilon_{\we{k}}^5 \right)}{9 \omega_{\we{k}}^5 \left(
		\omega_{\we{k}}^2 - 4 \varepsilon_{\we{k}}^2 \right)^2 \left(
		\omega_{\we{k}} + 4 \varepsilon_{\we{k}} \right)^2} :=
		g_{\we{k}\we{k}}^{(15)}.
        \end{equation*}
	$g_{\we{k}\we{k}}^{(15)} < 0 $ for all $k > 0$.
        \item If $ b > c$, $ b > e$, $ d < b \le 3d$ and $k \in \left[ \frac{b-d}{2a}
		, \frac{b+d}{4a} \right]$, then
        \begin{equation*}
                g_{\we{k}\we{k}}=
		-\frac{D_{\we{k}}^6 \left( 81 \omega_{\we{k}}^4 + 1320
		\omega_{\we{k}}^3 \varepsilon_{\we{k}} + 6544 \omega_{\we{k}}^2
		\varepsilon_{\we{k}}^2 + 13056 \omega_{\we{k}} \varepsilon_{\we{k}}^3
		+ 9216 \varepsilon_{\we{k}}^4 \right)}{9 \omega_{\we{k}}^5 \left(
		\omega_{\we{k}} + 4 \varepsilon_{\we{k}} \right)^2 \left(
		\omega_{\we{k}} + 2 \varepsilon_{\we{k}} \right)^2} < 0.
        \end{equation*}
        \item If $ b > c$, $ b > e$, $ d < b \le 3d$ and $k \in \left( \frac{b+d}{4a}
		, \frac{b+d}{2a} \right]$, then
        \begin{equation*}
		\begin{split}
                g_{\we{k}\we{k}} & =
		-\frac{D_{\we{k}}^6}{9 \omega_{\we{k}}^5 \left(
                \omega_{\we{k}}^2 - 16 \varepsilon_{\we{k}}^2 \right)^2 \left(
                \omega_{\we{k}} + 4 \varepsilon_{\we{k}} \right) \left( 
                \omega_{\we{k}} + 2 \varepsilon_{\we{k}} \right)^2}\\
		& \times \left( 9 \omega_{\we{k}}^7 + 36 \omega_{\we{k}}^6
		\varepsilon_{\we{k}} - 336 \omega_{\we{k}}^5 \varepsilon_{\we{k}}^2 -
		2560 \omega_{\we{k}}^4 \varepsilon_{\we{k}}^3 -3264 \omega_{\we{k}}^3
		\varepsilon_{\we{k}}^4 
		14592 \omega_{\we{k}}^2
		\varepsilon_{\we{k}}^5 + 46080 \omega_{\we{k}} \varepsilon_{\we{k}}^6
		+ 36864 \varepsilon_{\we{k}}^7 \right):=
		g_{\we{k}\we{k}}^{(16)}.\\
	\end{split}
        \end{equation*}
	$g_{\we{k}\we{k}}^{(16)} < 0$ iff $k \in \left( 0 , 0.142 b/a
	\right) \cup \left( 0.345 b/a , \infty \right)$.
        \item If $ b > c$, $ b > e$, $ d < b \le 3d$ and $k \in \left( \frac{b+d}{2a}
		, \infty \right)$, then $g_{\we{k}\we{k}}=
		g_{\we{k}\we{k}}^{(14)}$. 
        \item If $ b > c$, $ b > e$, $  b > 3d$ and $k \in \left( 0 , \frac{b-d}{4a}
		\right) \cup \left( \frac{b+d}{4a} , \frac{b-d}{2a} \right) \cup
		\left( \frac{b+d}{2a} , \infty \right)$, then $g_{\we{k}\we{k}}=
                g_{\we{k}\we{k}}^{(14)}$.
        \item If $ b > c$, $ b > e$, $  b > 3d$ and $k \in \left[ \frac{b-d}{4a} ,
		\frac{b+d}{4a} \right]$, then $g_{\we{k}\we{k}}=
                g_{\we{k}\we{k}}^{(15)}$.
        \item If $ b > c$, $ b > e$, $  b > 3d$ and $k \in \left[ \frac{b-d}{2a} ,
		\frac{b+d}{2a} \right]$, then $g_{\we{k}\we{k}}=
                g_{\we{k}\we{k}}^{(16)}$.	
\end{itemize}
\section{\label{b}Higher order terms of Fr\"ohlich's~transformation}
Evaluation of successive commutators of the~expansion~\eqref{app_h_trans} is 
a~time-consuming task. Below the~final results are given for the~relevant orders.

Since~$S=B-B^{\ast}$ in Eq.~\eqref{S}, all expansion terms have the~form~$g+g^{\ast}$. 
We focus on the~first summand. 

The following formulae for boson~operators
\begin{equation*}
        \left[b_1,b_2 b_3\right]=\left[ b_1,b_2\right]b_3+b_2\left[b_1,b_3\right],
\end{equation*}
\begin{equation*}
	\left[b_1 b_2,b_3 b_4\right]=b_1 b_3\left[b_2,b_4\right]+b_1\left[b_2,b_3\right] 
	b_4 + b_3\left[b_1,b_4\right]b_3+\left[b_1,b_3\right]b_4 b_2,
\end{equation*}
and fermion~operators
\begin{equation*}
        \left[f_1,f_2 f_3\right]=\left\{ f_1 , f_2 \right\}f_3-f_2\left\{ f_1 , f_3 
	\right\},
\end{equation*}
\begin{equation*}
        \left[f_1 f_2,f_3 f_4\right]=\left\{ f_1 , f_3 \right\}f_4 f_2 -f_3\left\{ f_1 , 
	f_4 \right\}f_2 + f_1\left\{ f_2 , f_3 \right\}f_4-f_1 f_3\left\{ f_2 , f_4 
	\right\},
\end{equation*}
will be used. In particular, for fermion creation and annihilation operators,
\begin{equation}
  	\left[a_k^\ast a_l,a_q^\ast a_r\right]=\delta_{lq}a_k^\ast a_r -\delta_{kr} 
	a_q^\ast a_l.
\label{com_4f}
\end{equation}
\subsection{Fourth order}
To calculate 4th~order terms, let us rewrite the 3rd~order expression~\eqref{3order} as 
a~sum of~$H_{\mathrm{int}}^{(3)}=\frac{1}{2}\left[S,\left[S,H_{\mathrm{int}}\right] 
\right]$ and~$H_0^{(3)}=-\frac{1}{6}\left[S,\left[S,\left[S,H_0\right]\right]\right]$:
\begin{equation*}
	H_{\mathrm{int},0}^{(3)}=\sum_{\we{q}\we{k}}A_{\we{k}\we{q}}^{int,0} 
	b_{\we{q}}^\ast n_{\we{k}} \gamma_{\we{q}}^\ast
	+\sum_{\we{q}\we{k}\we{w}\we{k'}}b_{\we{w}}^\ast 
	\left\{B_{\we{k}\we{q}\we{w}\we{k'}}^{int,0} a_{\we{k}-\we{w}}^\ast 
	a_{\we{k}-\we{q}}a_{\we{k'}-\we{q}}^\ast a_{\we{k'}}
	+ C_{\we{k}\we{q}\we{w}\we{k'}}^{int,0} a_{\we{k'}}^\ast 
	a_{\we{k'}-\we{q}}a_{\we{k}-\we{q}}^\ast a_{\we{k}+\we{w}} \right\}+c.c.,
\end{equation*}
with
\begin{equation*}
	A_{\we{k}\we{q}}^{\mathrm{int}}=\frac{1}{2}iD_{\we{q}}\bigl(\phi^\ast(\we{k},
	\we{q})+\phi(\we{k}+\we{q},\we{q})\bigr)+c.c.,
\end{equation*}
\begin{equation*}
	A_{\we{k}\we{q}}^{0}=\frac{1}{6}\left(\varepsilon_{\we{k}-\we{q}}-
	\varepsilon_{\we{k}}+\omega_{\we{q}} \right)|\phi(\we{k},\we{q})|^2
	-\frac{1}{6}\left(\varepsilon_{\we{k}}-\varepsilon_{\we{k}+\we{q}}+
	\omega_{\we{q}} \right) |\phi(\we{k}+\we{q},\we{q})|^2 +c.c.,
\end{equation*}
\begin{equation*}
	B_{\we{k}\we{q}\we{w}\we{k'}}^{\mathrm{int}}= 
	\frac{1}{2}iD_{\we{q}}\Bigl(\phi^\ast(\we{k'},\we{q})\phi^\ast(\we{k-\we{q}},
	\we{w})-\phi^\ast(\we{k'},\we{q})\phi^\ast(\we{k},\we{w})+
	\phi^\ast(\we{k},\we{w})\phi(\we{k},\we{q})
	-\phi^\ast(\we{k}-\we{q},\we{w})\phi(\we{k}-\we{w},\we{q})\Bigr),
\end{equation*}
\begin{equation*}
\begin{split}
     	B_{\we{k}\we{q}\we{w}\we{k'}}^{0} & = \frac{1}{6}\phi^\ast(\we{k'},\we{q}) 
	\Bigl\{\phi(\we{k},\we{q}) \phi^\ast(\we{k},\we{w})\left(\varepsilon_{\we{k}}-
	\varepsilon_{\we{k}-\we{q}}+\varepsilon_{\we{k'}-\we{q}}-
	\varepsilon_{\we{k'}}\right)\\
	& +\phi(\we{k}-\we{w},\we{q})\phi^\ast(\we{k}-\we{q},\we{w})\left( 
	\varepsilon_{\we{k}-\we{w}-\we{q}}-\varepsilon_{\we{k}-\we{w}}-
     	\varepsilon_{\we{k'}-\we{q}}+\varepsilon_{\we{k'}}\right)\Bigr\},\\
\end{split}
\end{equation*}
\begin{equation*}
\begin{split}
	C_{\we{k}\we{q}\we{w}\we{k'}}^{\mathrm{int}}&=\frac{1}{2}iD_{\we{q}}
	\Bigl(\phi(\we{k'},\we{q}) \phi^\ast(\we{k}+\we{w}-\we{q},\we{w})-\phi(\we{k'},
	\we{q})\phi^\ast(\we{k}+\we{w},\we{w})\\
	&-\phi^\ast(\we{k}+\we{w}-\we{q},\we{w})\phi^\ast(\we{k}+\we{w},\we{q})+
	\phi^\ast(\we{k}+\we{w},\we{w})\phi^\ast(\we{k},\we{q})\Bigr),\\
\end{split}
\end{equation*}
\begin{equation*}
\begin{split}
    	C_{\we{k}\we{q}\we{w}\we{k'}}^{0}&=\frac{1}{6}\phi(\we{k'},\we{q})\Bigl\{ 
	\phi^\ast(\we{k}+\we{w},\we{q}) \phi^\ast(\we{k}+\we{w}-\we{q},\we{w}) 
	\left(\varepsilon_{\we{k'}-\we{q}}-\varepsilon_{\we{k'}}-
    	\varepsilon_{\we{k}+\we{w}-\we{q}}+\varepsilon_{\we{k}+\we{w}}\right)\\
	&+\phi^\ast(\we{k},\we{q})\phi^\ast(\we{k}+\we{w},\we{w})\left( 
	\varepsilon_{\we{k}-\we{q}}-\varepsilon_{\we{k}}-\varepsilon_{\we{k'}-\we{q}}+
	\varepsilon_{\we{k'}}\right)\Bigr\}.\\
\end{split}
\end{equation*}

The 4th~order term in Eq.~\eqref{h_trans} equals
\begin{equation*}
	H^{(4)}=-\frac{1}{6}\left[S,\left[S,\left[S,H_{\mathrm{int}} \right]\right] 
	\right] + \frac{1}{24}\left[S,\left[S,\left[S,\left[S,H_0\right]\right]\right] 
	\right]
	=-\frac{1}{3}\left[S,H_{\mathrm{int}}^{(3)}\right]-\frac{1}{4}\left[S,H_0^{(3)} 
	\right]	:=H_{\mathrm{int}}^{(4)}+H_0^{(4)}.
\end{equation*}
Explicitly, 
\begin{equation}
\begin{split}
	&H_{\mathrm{int},0}^{(4)}=\sum_{\we{k}\we{q}}A_{\mathrm{int},0}^{(4)} 
	\gamma_{\we{q}}  n_{\we{k}} \gamma_{\we{q}}^\ast +
	\sum_{\we{k}\we{q}\we{k'}\we{w}}B_{\mathrm{int},0}^{(4)} \gamma_{\we{w}} 
	a_{\we{k}-\we{w}}^\ast a_{\we{k}-\we{q}}a_{\we{k'}-\we{q}}^\ast a_{\we{k'}}
	+\sum_{\we{k}\we{q}\we{k'}\we{w}}C_{\mathrm{int},0}^{(4)} \gamma_{\we{w}} 
	a_{\we{k'}}^\ast a_{\we{k'}-\we{q}}a_{\we{k}-\we{q}}^\ast a_{\we{k}+\we{w}}\\
	& + \sum_{\we{k}\we{q}\we{k'}} D_{\mathrm{int},0}^{(4)}b_{\we{q}}^\ast b_{\we{q}} 
	n_{\we{k}} n_{\we{k'}}
	+\sum_{\we{k}\we{q}}E_{\mathrm{int},0}^{(4)}b_{\we{q}}^\ast b_{\we{q}} 
	a_{\we{k}}^\ast a_{\we{k}-\we{q}}\gamma_{\we{q}}^\ast+
	\sum_{\we{k}\we{q}\we{k'}\we{w}} F_{\mathrm{int},0}^{(4)} b_{\we{w}}^\ast 
	b_{\we{w}}a_{\we{k}-\we{w}}^\ast a_{\we{k}-\we{q}}a_{\we{k'}-\we{q}}^\ast 
	a_{\we{k'}-\we{w}}+\\
	&+\sum_{\we{k}\we{q}\we{w}\we{k'}}G_{\mathrm{int},0}^{(4)}b_{\we{w}}^\ast 
	b_{\we{w}} a_{\we{k'}+\we{w}}^\ast a_{\we{k'}-\we{q}}a_{\we{k}-\we{q}}^\ast 
	a_{\we{k}+\we{w}} 
	+ \sum_{\we{k}\we{q}\we{w}\we{k'}} I_{\mathrm{int},0}^{(4)}b_{\we{w}}^\ast 
	b_{\we{w}}a_{\we{k}}^\ast a_{\we{k}-\we{q}}a_{\we{k'}-\we{q}}^\ast a_{\we{k'}}
	+c.c.,\\
\end{split}
\label{4order}
\end{equation}
where
\begin{equation*}
	A_{\mathrm{int}}^{(4)}=\frac{1}{3}A_{\we{k}\we{q}}^{\mathrm{int}}, \qquad 
	A_{0}^{(4)}= \frac{1}{4} A_{\we{k}\we{q}}^{0}, \qquad
	B_{\mathrm{int}}^{(4)}=\frac{1}{3}B_{\we{k}\we{q}\we{w}\we{k'}}^{\mathrm{int}}
	, \qquad B_{0}^{(4)}=\frac{1}{4}B_{\we{k}\we{q}\we{w}\we{k'}}^{0}.
\end{equation*}
Other coefficients in Eq.~\eqref{4order} arise according to the~same scheme, viz, 
\begin{equation*}
	C_{\mathrm{int}}^{(4)}=\frac{1}{3} C_{\we{k}\we{q}\we{w}\we{k'}}^{\mathrm{int}} 
	, \qquad 
	D_{\mathrm{int}}^{(4)}=\frac{1}{3}A_{\we{k}\we{q}}^{\mathrm{int}}
	\Bigl( |\phi(\we{k'},\we{q})|^2-	|\phi(\we{k'}+\we{q},\we{q})|^2\Bigr),
\end{equation*}
\begin{equation*}
	E_{\mathrm{int}}^{(4)}=\frac{1}{3}\left(A_{\we{k}-\we{q},\we{q}}^{\mathrm{int}}-
	A_{\we{k}\we{q}}^{\mathrm{int}}\right)\phi(\we{k},\we{q}), \qquad 
	F_{\mathrm{int}}^{(4)}=\frac{1}{3}\Bigl( B_{\we{k},\we{q},\we{w},\we{k'}-
	\we{w}}^{\mathrm{int}} \phi(\we{k'}-\we{q},\we{w})-
	B_{\we{k},\we{q},\we{w},\we{k'}}^{\mathrm{int}}\phi(\we{k'},\we{w}) 
	\Bigr),
\end{equation*}
\begin{equation*}
	G_{\mathrm{int}}^{(4)}=\frac{1}{3}\Bigl( C_{\we{k},\we{q},\we{w},
	\we{k'}}^{\mathrm{int}} \phi(\we{k'}+ \we{w},\we{w})-C_{\we{k},\we{q},\we{w},
	\we{k'}+\we{w}}^{\mathrm{int}}\phi(\we{k'}+\we{w}-\we{q},\we{w})\Bigr),
\end{equation*}
\begin{equation*}
		I_{\mathrm{int}}^{(4)}=\frac{1}{3}\Bigl( B_{\we{k},\we{q},\we{w},
		\we{k'}}^{\mathrm{int}}\phi(\we{k},\we{w})-B_{\we{k}+\we{w},\we{q},
		\we{w},\we{k'}}^{\mathrm{int}}\phi(\we{k}+\we{w}-\we{q},\we{w})
		 + C_{\we{k'}-\we{w},\we{q},\we{w},\we{k}}^{\mathrm{int}}\phi(\we{k'}-
		\we{q},\we{w}) - C_{\we{k'},\we{q},\we{w},\we{k}}^{\mathrm{int}} 
		\phi(\we{k'}+\we{w},\we{w})\Bigr).
\end{equation*}
and similarly for~$C_0^{(4)}$, $D_0^{(4)}$,~etc.
\subsection{Fifth order}
Proceeding analogously as with 4th~order, one obtains 
\begin{equation}
	H^{(5)}=\frac{1}{24}\left[S,\left[S,\left[S,\left[S,H_{\mathrm{int}} \right] 
	\right]\right]\right] - \frac{1}{120}\left[S,\left[S,\left[S,\left[S,\left[S,
	H_{0}\right] \right]\right] \right] \right]
	=-\frac{1}{4}\left[S,H_{\mathrm{int}}^{(4)}\right]-\frac{1}{5}\left[S,H_0^{(4)} 
	\right] := H_{\mathrm{int}}^{(5)}+H_0^{(5)},
\label{5order}
\end{equation}
\begin{equation*}
\begin{split}
	&H_{\mathrm{int},0}^{(5)}=\sum_{\we{q}\we{k}\we{w}\we{k'}}A_{int,0}^{(5)} 
	b_{\we{w}}^\ast \gamma_{\we{q}}n_{\we{k}}a_{\we{k'}-\we{w}}^\ast a_{\we{k'}+
	\we{q}}+
	\sum_{\we{q}\we{k}\we{w}}B_{\mathrm{int},0}^{(5)}b_{\we{w}}^\ast \gamma_{\we{q}} 
	a_{\we{k}}^\ast a_{\we{k}-\we{w}}\gamma_{\we{q}}^\ast
	+\sum_{\we{q}\we{k}\we{w}\we{k'}}C_{\mathrm{int},0}^{(5)}b_{\we{w}}^\ast 
	a_{\we{k'}-\we{w}} a_{\we{k'}-\we{q}}n_{\we{k}}\gamma_{\we{q}}^\ast\\
	& + \sum_{\we{q}\we{k}\we{w}\we{k'}\we{u}}D_{\mathrm{int},0}^{(5)}b_{\we{u}}^\ast 
	\gamma_{\we{w}}a_{\we{k}-\we{w}}^\ast a_{\we{k}-\we{q}}a_{\we{k'}-\we{q}}^\ast 
	a_{\we{k'}+\we{u}}
	+\sum_{\we{q}\we{k}\we{w}\we{k'}\we{u}} E_{\mathrm{int},0}^{(5)} b_{\we{u}}^\ast 
	\gamma_{\we{w}} a_{\we{k}-\we{w}-\we{u}}^\ast a_{\we{k}-\we{q}}
	a_{\we{k'}-\we{q}}^\ast a_{\we{k'}}\\
	&+\sum_{\we{q}\we{k}\we{w}\we{k'}\we{u}\we{l}}F_{\mathrm{int},0}^{(5)} 
	b_{\we{u}}^\ast a_{\we{l}-\we{u}}^\ast a_{\we{l}-\we{w}}a_{\we{k}-\we{w}}^\ast 
	a_{\we{k}-\we{q}}a_{\we{k'}-\we{q}}^\ast a_{\we{k'}}
	+\sum_{\we{q}\we{k}\we{w}\we{k'}\we{u}}G_{\mathrm{int},0}^{(5)}b_{\we{u}} 
	\gamma_{\we{w}}a_{\we{k}-\we{w}}^\ast a_{\we{k}-\we{q}}a_{\we{k'}-\we{q}}^\ast 
	a_{\we{k'}-\we{u}}\\
	&+\sum_{\we{q}\we{k}\we{w}\we{k'}\we{u}}I_{\mathrm{int},0}^{(5)}b_{\we{u}} 
	\gamma_{\we{w}}a_{\we{k}-\we{w}}^\ast a_{\we{k}-\we{q}-\we{u}}
	a_{\we{k'}-\we{q}}^\ast a_{\we{k'}}
	+\sum_{\we{q}\we{k}\we{w}\we{k'}\we{u}\we{l}} J_{\mathrm{int},0}^{(5)}b_{\we{u}} 
	a_{\we{l}+\we{w}}^\ast a_{\we{l}-\we{u}}a_{\we{k}-\we{w}}^\ast a_{\we{k}-\we{q}} 
	a_{\we{k'}-\we{q}}^\ast a_{\we{k'}}\\
   	& +\sum_{\we{q}\we{k}\we{w}\we{k'}\we{u}}K_{\mathrm{int},0}^{(5)} b_{\we{u}}^\ast 
	\gamma_{\we{w}} a_{\we{k'}}^\ast a_{\we{k'}-\we{q}}a_{\we{k}-\we{q}-\we{u}}^\ast 
	a_{\we{k}+\we{w}}
	+\sum_{\we{q}\we{k}\we{w}\we{k'}\we{u}} L_{\mathrm{int},0}^{(5)} b_{\we{u}}^\ast 
	\gamma_{\we{w}} a_{\we{k'}-\we{u}}^\ast a_{\we{k'}-\we{q}}a_{\we{k}-\we{q}}^\ast 
	a_{\we{k}+\we{w}}\\
	& + \sum_{\we{q}\we{k}\we{w}\we{k'}\we{u}\we{l}}M_{\mathrm{int},0}^{(5)} 
	b_{\we{u}}^\ast a_{\we{l}-\we{u}}^\ast a_{\we{l}-\we{q}}a_{\we{k'}}^\ast 
	a_{\we{k'}-\we{q}}a_{\we{k}-\we{q}}^\ast a_{\we{k'}+\we{w}}
	+\sum_{\we{q}\we{k}\we{w}\we{k'}\we{u}}N_{\mathrm{int},0}^{(5)} b_{\we{u}} 
	\gamma_{\we{w}} a_{\we{k'}}^\ast a_{\we{k'}-\we{q}}a_{\we{k}-\we{q}}^\ast 
	a_{\we{k}+\we{w}-\we{u}}\\
   	& +\sum_{\we{q}\we{k}\we{w}\we{k'}\we{u}}O_{\mathrm{int},0}^{(5)}b_{\we{u}} 
	\gamma_{\we{w}}a_{\we{k'}+\we{u}}^\ast a_{\we{k'}-\we{q}}a_{\we{k}-\we{q}}^\ast 
	a_{\we{k}+\we{w}}
	+\sum_{\we{q}\we{k}\we{w}\we{k'}\we{u}\we{l}} P_{\mathrm{int},0}^{(5)}b_{\we{u}} 
	a_{\we{l}+\we{w}}^\ast a_{\we{l}-\we{u}}a_{\we{k'}}^\ast a_{\we{k'}-\we{q}} 
	a_{\we{k}-\we{q}}^\ast a_{\we{k}+\we{w}}
   	+\sum_{\we{q}\we{k}\we{k'}}R_{\mathrm{int},0}^{(5)}b_{\we{q}}^\ast n_{\we{k}} 
	n_{\we{k'}} \gamma_{\we{q}}^\ast\\
	&+\sum_{\we{q}\we{k}}S_{\mathrm{int},0}^{(5)}b_{\we{q}}^\ast a_{\we{k}}^\ast 
	a_{\we{k}-\we{q}} \gamma_{\we{q}}^\ast \gamma_{\we{q}}^\ast
	+\sum_{\we{q}\we{k}}S_{\mathrm{int},0}^{(5)}b_{\we{q}}\gamma_{\we{q}} 
	a_{\we{k}}^\ast a_{\we{k}-\we{q}}\gamma_{\we{q}}^\ast
   	+\sum_{\we{q}\we{k}\we{w}\we{k'}}T_{\mathrm{int},0}^{(5)}b_{\we{w}}^\ast 
	a_{\we{k}-\we{w}}^\ast a_{\we{k}-\we{q}} a_{\we{k'}-\we{q}}^\ast 
	a_{\we{k'}-\we{w}} \gamma_{\we{w}}^\ast\\
	&+\sum_{\we{q}\we{k}\we{w}\we{k'}} U_{\mathrm{int},0}^{(5)}b_{\we{w}}^\ast 
	a_{\we{k'}+\we{w}}^\ast a_{\we{k'}-\we{q}}a_{\we{k}-\we{q}}^\ast 
	a_{\we{k}+\we{w}} \gamma_{\we{w}}^\ast 
	+ \sum_{\we{q}\we{k}\we{w}\we{k'}}W_{\mathrm{int},0}^{(5)}b_{\we{w}}^\ast 
	a_{\we{k}}^\ast a_{\we{k}-\we{q}}a_{\we{k'}-\we{q}}^\ast a_{\we{k'}} 
	\gamma_{\we{w}}^\ast+c.c.,\\
\end{split}
\end{equation*}								   
where
\begin{equation*}
	A_{\mathrm{int},0}^{(5)}=-\left\{ \frac{1}{4} , \frac{1}{5}\right\}\left( 
	A_{\mathrm{int},0}^{(4)}+A_{\mathrm{int},0}^{(4)\ast}\right)
	\times\Bigl( \phi^\ast(\we{k'},\we{w})\phi^\ast(\we{k'}+\we{q},\we{q}) - 
	\phi^\ast(\we{k'}+\we{q},\we{w})\phi^\ast(\we{k'}+\we{q}-\we{w},\we{q}) 
	\Bigr),
\end{equation*}
\begin{equation*}
	B_{\mathrm{int},0}^{(5)}=-\left\{ \frac{1}{4} , \frac{1}{5}\right\}\left( 
	A_{\mathrm{int},0}^{(4)}+A_{\mathrm{int},0}^{(4)\ast}-
	A_{\mathrm{int},0}^{(4)\we{k}\to\we{k}-\we{w}}-A_{\mathrm{int},
	0}^{(4)\ast\we{k}\to\we{k}-\we{w}}\right)\phi^\ast(\we{k},\we{w}),
\end{equation*}
\begin{equation*}
	C_{\mathrm{int},0}^{(5)}=-\left\{ \frac{1}{4} , \frac{1}{5}\right\}\left( 
	A_{\mathrm{int},0}^{(4)}+A_{\mathrm{int},0}^{(4)\ast} \right) 
	\Bigl( \phi^\ast(\we{k'},\we{w})\phi^\ast(\we{k'},\we{q})-
	\phi^\ast(\we{k'}-\we{q},\we{w})\phi^\ast(\we{k'}-\we{w},\we{q})\Bigr),
\end{equation*}
\begin{equation*}
	D_{\mathrm{int},0}^{(5)}=-\left\{ \frac{1}{4} , \frac{1}{5}\right\}\left( 
	B_{\mathrm{int},0}^{(4)\we{k'}\to\we{k'}+\we{u}}\phi^\ast(\we{k'}+\we{u}-\we{q},
	\we{u})-B_{\mathrm{int},0}^{(4)} \phi^\ast(\we{k'}+\we{u},\we{u})\right),
\end{equation*}
\begin{equation*}
	E_{\mathrm{int},0}^{(5)}=-\left\{ \frac{1}{4} , \frac{1}{5}\right\}\left( 
	B_{\mathrm{int},0}^{(4)} \phi^\ast(\we{k}-\we{w},\we{u})-
	B_{\mathrm{int},0}^{(4)\we{k}\to\we{k}-\we{u}} \phi^\ast(\we{k}-\we{q},\we{u}) 
	\right),
\end{equation*}
\begin{equation*}
	F_{\mathrm{int},0}^{(5)}=-\left\{\frac{1}{4} , \frac{1}{5}\right\}
	B_{\mathrm{int},0}^{(4)} \Bigl( \phi^\ast(\we{l},\we{u})\phi(\we{l},\we{w})-
	\phi^\ast(\we{l}-\we{w},\we{u})\phi(\we{l}-\we{u},\we{w})\Bigr),
\end{equation*}
\begin{equation*}
	G_{\mathrm{int},0}^{(5)}=\left\{\frac{1}{4} ,\frac{1}{5}\right\} 
	\left(B_{\mathrm{int},0}^{(4)\we{k'}\to\we{k'}-\we{u}}\phi(\we{k'}-\we{q},\we{u})
	-B_{\mathrm{int},0}^{(4)}\phi(\we{k'},\we{u})\right),
\end{equation*}
\begin{equation*}
	I_{\mathrm{int},0}^{(5)}=\left\{\frac{1}{4} , \frac{1}{5}\right\}
	\left(B_{\mathrm{int},0}^{(4)\we{k}\to\we{k}-\we{u}}\phi(\we{k}-\we{w},\we{u})
	-B_{\mathrm{int},0}^{(4)}\phi(\we{k}-\we{q},\we{u}) \right),
\end{equation*}
\begin{equation*}
	J_{\mathrm{int},0}^{(5)}=\left\{\frac{1}{4} , \frac{1}{5}\right\}
	B_{\mathrm{int},0}^{(4)} \Bigl( \phi(\we{l}+\we{w},\we{u})\phi(\we{l}-\we{u}+
	\we{w},\we{w})-\phi(\we{l},\we{u})\phi(\we{l}+\we{w},\we{w})\Bigr),
\end{equation*}
\begin{equation*}
	K_{\mathrm{int},0}^{(5)}=-\left\{\frac{1}{4} ,\frac{1}{5}\right\}
	\left(C_{\mathrm{int},0}^{(4)} \phi^\ast(\we{k}-\we{q},\we{u})-
	C_{\mathrm{int},0}^{(4)\we{k}\to\we{k}-\we{u}}\phi^\ast(\we{k}+\we{w},\we{u}) 
	\right),
\end{equation*}
\begin{equation*}
	L_{\mathrm{int},0}^{(5)}=-\left\{\frac{1}{4} ,\frac{1}{5}\right\}
	\left(C_{\mathrm{int},0}^{(4)} \phi^\ast(\we{k'},\we{u})-
	C_{\mathrm{int},0}^{(4)\we{k'}\to\we{k'}-\we{u}}\phi^\ast(\we{k'}-\we{q},\we{u}) 
	\right),
\end{equation*}
\begin{equation*}
	M_{\mathrm{int},0}^{(5)}=-\left\{\frac{1}{4} , \frac{1}{5}\right\}
	C_{\mathrm{int},0}^{(4)} \Bigl( \phi^\ast(\we{l},\we{u})\phi(\we{l},\we{w})-
	\phi^\ast(\we{l}-\we{w},\we{u})\phi(\we{l}-\we{u},\we{w})\Bigr),
\end{equation*}
\begin{equation*}
	N_{\mathrm{int},0}^{(5)}=\left\{\frac{1}{4} ,\frac{1}{5}\right\}
	\left(C_{\mathrm{int},0}^{(4)\we{k}\to\we{k}-\we{u}}\phi(\we{k}-\we{q},\we{u})
	-C_{\mathrm{int},0}^{(4)}\phi(\we{k}+\we{w},\we{u})\right),
\end{equation*}
\begin{equation*}
	O_{\mathrm{int},0}^{(5)}=\left\{\frac{1}{4} ,\frac{1}{5}\right\}
	\left(C_{\mathrm{int},0}^{(4)} \phi(\we{k'}+\we{u},\we{u})-
	C_{\mathrm{int},0}^{(4)\we{k'}\to\we{k'}+\we{u}}\phi(\we{k'}+\we{u}-\we{q},
	\we{u}) \right),
\end{equation*}
\begin{equation*}
	P_{\mathrm{int},0}^{(5)}=\left\{\frac{1}{4} , \frac{1}{5}\right\}
	C_{\mathrm{int},0}^{(4)} \Bigl( \phi(\we{l}+\we{w},\we{u})\phi(\we{l}-\we{u}+
	\we{w},\we{w})-\phi(\we{l},\we{u})\phi(\we{l}+\we{w},\we{w})\Bigr),
\end{equation*}
\begin{equation*}
	R_{\mathrm{int},0}^{(5)}=\left\{\frac{1}{4} , \frac{1}{5}\right\}
	\left(D_{\mathrm{int},0}^{(4)} + D_{\mathrm{int},0}^{(4)\ast}\right),\qquad
	S_{\mathrm{int},0}^{(5)}=\left\{\frac{1}{4} , \frac{1}{5}\right\}
	E_{\mathrm{int},0}^{(4)}, \qquad
	T_{\mathrm{int},0}^{(5)}=\left\{\frac{1}{4} , \frac{1}{5}\right\} 
	\left(F_{\mathrm{int},0}^{(4)}+ F_{\mathrm{int},0}^{(4)\ast\we{k} 
	\leftrightarrows\we{k'}}\right), 
\end{equation*}
\begin{equation*}
	U_{\mathrm{int},0}^{(5)}=\left\{\frac{1}{4} ,\frac{1}{5}\right\}
	\left(G_{\mathrm{int},0}^{(4)}+ G_{\mathrm{int},0}^{(4)\ast\we{k} 
	\leftrightarrows\we{k'}}\right), \qquad
	W_{\mathrm{int},0}^{(5)}=\left\{\frac{1}{4} , \frac{1}{5}\right\} 
	\left( I_{\mathrm{int},0}^{(4)}+ I_{\mathrm{int},0}^{(4)\ast\we{k} 
	\leftrightarrows\we{k'}}\right).
\end{equation*}
$1/4$ in the~curly brackets refering to int~terms and~$1/5$ to $0$~terms.
\subsection{Sixth order}
\begin{equation}
	H^{(6)}=-\frac{1}{120}\left[S,\left[S,\left[S,\left[S,\left[S,H_{\mathrm{int}} 
	\right]\right]\right]\right]\right] + \frac{1}{720}\left[S,\left[S,\left[S,
	\left[S,\left[S,\left[S,H_{0}\right]\right] \right]\right]\right]\right]
	=-\frac{1}{5}\left[S,H_{\mathrm{int}}^{(5)}\right] - \frac{1}{6}\left[S,
	H_0^{(5)}\right] := H_{\mathrm{int}}^{(6)}+H_0^{(6)},
\label{6order}
\end{equation}
\begin{equation}
\begin{split}
	&H_{\mathrm{int},0}^{(6)}=\left\{\frac{1}{5} , \frac{1}{6}\right\} \left\{ 
	\sum_{\we{q}\we{k}\we{w}\we{k'}} A_{\mathrm{int},0}^{(5)} \gamma_{\we{w}} 
	\gamma_{\we{q}} n_{\we{k}}a_{\we{k'}-\we{w}}^\ast a_{\we{k'}+\we{q}}
	+\sum_{\we{q}\we{k}\we{w}}B_{\mathrm{int},0}^{(5)}\gamma_{\we{w}} 
	\gamma_{\we{q}}a_{\we{k}}^\ast a_{\we{k}-\we{w}}\gamma_{\we{q}}^\ast
	+\sum_{\we{q}\we{k}\we{w}\we{k'}}C_{\mathrm{int},0}^{(5)}\gamma_{\we{w}} 
	a_{\we{k'}-\we{w}} a_{\we{k'}-\we{q}}n_{\we{k}}\gamma_{\we{q}}^\ast\right.\\
	&+\sum_{\we{q}\we{k}\we{w}\we{k'}\we{u}}D_{\mathrm{int},0}^{(5)} \gamma_{\we{u}} 
	\gamma_{\we{w}} a_{\we{k}-\we{w}}^\ast a_{\we{k}-\we{q}}a_{\we{k'}-\we{q}}^\ast 
	a_{\we{k'}+\we{u}}
	+\sum_{\we{q}\we{k}\we{w}\we{k'}\we{u}} E_{\mathrm{int},0}^{(5)} \gamma_{\we{u}} 
	\gamma_{\we{w}} a_{\we{k}-\we{w}-\we{u}}^\ast a_{\we{k}-\we{q}}
	a_{\we{k'}-\we{q}}^\ast a_{\we{k'}}\\
	&+\sum_{\we{q}\we{k}\we{w}\we{k'}\we{u}\we{l}}F_{\mathrm{int},0}^{(5)} 
	\gamma_{\we{u}} a_{\we{l}-\we{u}}^\ast a_{\we{l}-\we{w}}a_{\we{k}-\we{w}}^\ast 
	a_{\we{k}-\we{q}}a_{\we{k'}-\we{q}}^\ast a_{\we{k'}}
	+\sum_{\we{q}\we{k}\we{w}\we{k'}\we{u}}G_{\mathrm{int},0}^{(5)}\gamma_{\we{w}} 
	a_{\we{k}-\we{w}}^\ast a_{\we{k}-\we{q}}a_{\we{k'}-\we{q}}^\ast 
	a_{\we{k'}-\we{u}} \gamma_{\we{u}}^\ast \\
  	& +\sum_{\we{q}\we{k}\we{w}\we{k'}\we{u}}I_{\mathrm{int},0}^{(5)}\gamma_{\we{w}} 
	a_{\we{k}-\we{w}}^\ast a_{\we{k}-\we{q}-\we{u}}a_{\we{k'}-\we{q}}^\ast 
	a_{\we{k'}}\gamma_{\we{u}}^\ast
	+\sum_{\we{q}\we{k}\we{w}\we{k'}\we{u}\we{l}}J_{\mathrm{int},0}^{(5)} 
	a_{\we{l}+\we{w}}^\ast a_{\we{l}-\we{u}}a_{\we{k}-\we{w}}^\ast 
	a_{\we{k}-\we{q}}a_{\we{k'}-\we{q}}^\ast a_{\we{k'}} \gamma_{\we{u}}^\ast\\
  	& +\sum_{\we{q}\we{k}\we{w}\we{k'}\we{u}}K_{\mathrm{int},0}^{(5)} \gamma_{\we{u}} 
	\gamma_{\we{w}} a_{\we{k'}}^\ast a_{\we{k'}-\we{q}}a_{\we{k}-\we{q}-\we{u}}^\ast 
	a_{\we{k}+\we{w}}
	+\sum_{\we{q}\we{k}\we{w}\we{k'}\we{u}} L_{\mathrm{int},0}^{(5)} \gamma_{\we{u}} 
	\gamma_{\we{w}} a_{\we{k'}-\we{u}}^\ast a_{\we{k'}-\we{q}}a_{\we{k}-\we{q}}^\ast 
	a_{\we{k}+\we{w}}\\
  	& +\sum_{\we{q}\we{k}\we{w}\we{k'}\we{u}\we{l}}M_{\mathrm{int},0}^{(5)} 
	\gamma_{\we{u}} a_{\we{l}-\we{u}}^\ast a_{\we{l}-\we{q}}a_{\we{k'}}^\ast 
	a_{\we{k'}-\we{q}}a_{\we{k}-\we{q}}^\ast a_{\we{k'}+\we{w}}
	+\sum_{\we{q}\we{k}\we{w}\we{k'}\we{u}}N_{\mathrm{int},0}^{(5)}\gamma_{\we{w}} 
	a_{\we{k'}}^\ast a_{\we{k'}-\we{q}}a_{\we{k}-\we{q}}^\ast 
	a_{\we{k}+\we{w}-\we{u}}\gamma_{\we{u}}^\ast \\
	& + \sum_{\we{q}\we{k}\we{w}\we{k'}\we{u}}O_{\mathrm{int},0}^{(5)}\gamma_{\we{w}} 
	a_{\we{k'}+\we{u}}^\ast a_{\we{k'}-\we{q}}a_{\we{k}-\we{q}}^\ast 
	a_{\we{k}+\we{w}}\gamma_{\we{u}}^\ast
	+\sum_{\we{q}\we{k}\we{w}\we{k'}\we{u}\we{l}}P_{\mathrm{int},0}^{(5)} 
	a_{\we{l}+\we{w}}^\ast a_{\we{l}-\we{u}}a_{\we{k'}}^\ast a_{\we{k'}-\we{q}} 
	a_{\we{k}-\we{q}}^\ast a_{\we{k}+\we{w}} \gamma_{\we{u}}^\ast
   	+\sum_{\we{q}\we{k}\we{k'}}R_{\mathrm{int},0}^{(5)} \gamma_{\we{q}} n_{\we{k}} 
	n_{\we{k'}} \gamma_{\we{q}}^\ast\\
	&+\sum_{\we{q}\we{k}}S_{\mathrm{int},0}^{(5)}\gamma_{\we{q}}a_{\we{k}}^\ast 
	a_{\we{k}-\we{q}} \gamma_{\we{q}}^\ast\gamma_{\we{q}}^\ast
	+\sum_{\we{q}\we{k}}S_{\mathrm{int},0}^{(5)}\gamma_{\we{q}}a_{\we{k}}^\ast 
	a_{\we{k}-\we{q}} \gamma_{\we{q}}^\ast \gamma_{\we{q}}^\ast
   	+\sum_{\we{q}\we{k}\we{w}\we{k'}}T_{\mathrm{int},0}^{(5)}\gamma_{\we{w}} 
	a_{\we{k}-\we{w}}^\ast a_{\we{k}-\we{q}} a_{\we{k'}-\we{q}}^\ast 
	a_{\we{k'}-\we{w}} \gamma_{\we{w}}^\ast \\
	& +\sum_{\we{q}\we{k}\we{w}\we{k'}}U_{\mathrm{int},0}^{(5)}\gamma_{\we{w}} 
	a_{\we{k'}+\we{w}}^\ast a_{\we{k'}-\we{q}}a_{\we{k}-\we{q}}^\ast 
	a_{\we{k}+\we{w}}\gamma_{\we{w}}^\ast
   	\left.+\sum_{\we{q}\we{k}\we{w}\we{k'}}W_{\mathrm{int},0}^{(5)}\gamma_{\we{w}} 
	a_{\we{k}}^\ast a_{\we{k}-\we{q}}a_{\we{k'}-\we{q}}^\ast a_{\we{k'}} 
	\gamma_{\we{w}}^\ast+ b^\ast b \left\{ \right\} +c.c.\right\}.\\
\end{split}
\label{fr_6ord}
\end{equation}
The~last term denoting all~$(6,2)$ expressions which are irrelevant after averaging over 
phonon vacuum. The~coefficients are the~same as in the 5th~order terms, except 
for~the factors~$1/5$,~$1/6$.
\begin{acknowledgments}
The~authors would like to thank \mbox{prof.\ J.\ Ma\'ckowiak} for valuable comments and 
suggestions.
\end{acknowledgments}
\bibliography{arts,books,misc}
\end{document}